\documentclass[final,leqno,onefignum,onetabnum]{siamltex}

\newcommand{\fontDiscrete}{\mathcal}
\newcommand{\dummySymb}{v}

\newcommand{\paramSymb}{\mu}
\newcommand{\solSymb}{x}
\newcommand{\sol}{\boldsymbol \solSymb}

\newcommand{\param}{\boldsymbol \paramSymb}
\newcommand{\paramDomain}{\mathcal D}
\newcommand{\solArg}[1]{\sol_{#1}}

\newcommand{\RR}[1]{\ensuremath{\mathbb{R}^{ #1 }}}
\newcommand{\RRplus}[1]{\ensuremath{\mathbb{R}_+^{ #1 }}}
\newcommand{\nbig}{N}
\newcommand{\spaceSymb}{s}
\newcommand{\nspacedof}{\nbig_\spaceSymb}

\newcommand{\nsmall}{n}
\newcommand{\nparam}{\nsmall_\mu}
\newcommand{\spatialSubspace}{\fontDiscrete S}
\newcommand{\defeq}{:=}
\newcommand{\scaledDecoder}{\boldsymbol{g}}
\newcommand{\scaledEncoder}{\boldsymbol{h}}
\newcommand{\dummy}{\boldsymbol {\dummySymb}}
\newcommand{\reddummy}{\hat{\dummy}}
\newcommand{\nbasisspace}{{\nsmall_\spaceSymb}}
\newcommand{\solapprox}{\tilde\sol}
\newcommand{\redsolapprox}{\hat\sol}
\newcommand{\redsolapproxArg}[1]{\redsolapprox_{#1}}
\newcommand{\basismatspaceSymb}{\Phi}
\newcommand{\basismatspace}{\boldsymbol{\basismatspaceSymb}}
\newcommand{\basisvecspaceSymb}{\phi}
\newcommand{\basisvecspace}{\boldsymbol{\basisvecspaceSymb}}

\newcommand{\snapshotSymb}{X}
\newcommand{\snapshots}{\boldsymbol \snapshotSymb}
\newcommand{\bmat}[1]{\begin{bmatrix}#1\end{bmatrix}}
\newcommand{\timeSymb}{t}
\newcommand{\ntimedof}{{\nbig_\timeSymb}}
\newcommand{\NN}{\mathbb{N}}
\newcommand{\natNo}{\NN}
\newcommand{\nat}[1]{\natNo(#1)}
\newcommand{\innat}[1]{\in\nat{#1}}
\newcommand{\ndof}{\nspacedof}
\def\Ubold{\boldsymbol{U}}
\def\Vbold{\boldsymbol{V}}
\def\Sigmabold{\boldsymbol{\Sigma}}
\def\ubold{\boldsymbol{u}}

\newcommand{\argmin}[1]{\underset{#1}{\text{argmin}}}
\newcommand{\samplematSymb}{Z}
\newcommand{\samplematNT}{\boldsymbol \samplematSymb}
\newcommand{\samplemat}{\samplematNT^T}
\newcommand{\bds}[1]{{\boldsymbol{#1}}}
\newcommand{\unitvec}{\bds{e}}
\newcommand{\unitvecArg}[1]{\unitvec_{#1}}

\newcommand{\resSymb}{r}
\newcommand{\nbasisres}{\nsmall_\resSymb}
\newcommand{\identity}[1]{\boldsymbol I_{#1}}
\newcommand{\res}{\boldsymbol \resSymb}
\newcommand{\redres}{\hat{\res}}

\newcommand{\basismatres}{\basismatspace_\resSymb}

\usepackage{fullpage}
\usepackage{upgreek}
\usepackage[mathscr]{eucal}
\usepackage{color}
\usepackage[usenames,dvipsnames,svgnames,table]{xcolor}
\usepackage{amsmath,amssymb,amsopn,mathtools}
\usepackage{graphicx}
\usepackage{hyperref}
\usepackage{relsize}
\usepackage{enumitem}
\usepackage[normalem]{ulem}
\usepackage{verbatim}
\usepackage{footmisc}
\usepackage{soul}
\usepackage{subfigure, algorithmic, algorithm}
\usepackage{epsfig}
\usepackage{multirow, booktabs}
\usepackage{siunitx}
\mathtoolsset{showonlyrefs=false}
\usepackage{xr}
\usepackage{morefloats}
\usepackage{float}
\usepackage{lineno}
\usepackage{nomencl}

\makenomenclature

\makeatletter

\hypersetup{
    colorlinks=true,   
    citecolor=blue,    
    filecolor=blue,    
    linkcolor=red,     
    urlcolor=blue}     

\numberwithin{equation}{section}

\newlist{Assumption}{enumerate}{1}
\setlist[Assumption]{label=A\arabic*}
\newlist{steps}{enumerate}{1}
\setlist[steps, 1]{label = Step \arabic*:}

\title{Gappy AE: A Nonlinear Approach for Gappy Data Reconstruction using Auto-Encoder}
\author{
    Youngkyu Kim\thanks{Center for Artificial Intelligence, 
    Korea Institute of Science and Technology, 
    Seongbuk-gu, Seoul 02792 Republic of Korea (youngkyu.kim@kist.re.kr, yoo@byoo.org)}
    \and
    Youngsoo Choi\thanks{Center for Applied Scientific Computing,
    Lawrence Livermore National Laboratory, 
    Livermore, CA 94550 USA (choi15@llnl.gov)}
    \and
    Byounghyun Yoo\footnotemark[1] $^{,}$\thanks{Correspondence}
}

\begin{document}

\setlength{\abovedisplayskip}{3pt}
\setlength{\belowdisplayskip}{3pt} 
\setlength{\abovedisplayshortskip}{3pt} 
\setlength{\belowdisplayshortskip}{3pt}

\maketitle

\begin{abstract}
 We introduce a novel data reconstruction algorithm known as Gappy auto-encoder (Gappy AE) to address the limitations associated with Gappy proper orthogonal decomposition (Gappy POD), a widely used method for data reconstruction when dealing with sparse measurements or missing data. Gappy POD has inherent constraints in accurately representing solutions characterized by slowly decaying Kolmogorov N-widths, primarily due to its reliance on linear subspaces for data prediction. In contrast, Gappy AE leverages the power of nonlinear manifold representations to address data reconstruction challenges of conventional Gappy POD. It excels at real-time state prediction in scenarios where only sparsely measured data is available, filling in the gaps effectively. This capability makes Gappy AE particularly valuable, such as for digital twin and image correction applications. To demonstrate the superior data reconstruction performance of Gappy AE with sparse measurements, we provide several numerical examples, including scenarios like 2D diffusion, 2D radial advection, and 2D wave equation problems. Additionally, we assess the impact of four distinct sampling algorithms - discrete empirical interpolation method, the S-OPT algorithm, Latin hypercube sampling, and uniformly distributed sampling - on data reconstruction accuracy. Our findings conclusively show that Gappy AE outperforms Gappy POD in data reconstruction when sparse measurements are given.
\end{abstract}


\section{Introduction}\label{sec:intro}
In digital twin applications, adopting efficient data-driven models for representing physical fields has become increasingly prevalent. To enhance efficiency, these models involve mapping from a low-dimensional to full-dimensional space. In other words, these learned models approximate solutions in the full space based on information from a reduced-dimensional space. This approach encompasses two distinct solution representations: linear subspaces and nonlinear manifolds. However, the key challenge is identifying a suitable generalized coordinate within the low-dimensional space.

Methods to find a suitable generalized coordinate can be categorized into two approaches: intrusive and non-intrusive. In intrusive methods, the governing equations are solved with a substituted solution approximation. Projection-based reduced-order models, such as the linear subspace reduced-order model \cite{carlberg2018conservative,choi2019accelerating,choi2020gradient,choi2020sns,copeland2022reduced,cheung2023local}, nonlinear manifold reduced-order model \cite{lee2020model,kim2022fast,kim2020efficient,diaz2023fast}, component-wise reduced-order model \cite{mcbane2021component,mcbane2022stress} and space-time reduced-order model \cite{choi2019space,choi2020space,kim2021efficient,taddei2021space}, are examples of intrusive methods. These techniques are widely employed to expedite physical simulations. Recent research in reduced order models (ROMs) focuses on enhancing the representability of autoencoders \cite{ romor2023non} and the computational efficiency of autoencoder training \cite{ cocola2023hyper, diaz2023fast}. In \cite{cocola2023hyper}, the authors reduced computational costs by training autoencoders on subsampled points only. During the online phase, the decoder outputs data corresponding to subsampled points, and then a POD basis matrix is used to reconstruct the data. The authors in \cite{diaz2023fast} present the domain decomposition (DD) method for nonlinear-manifold reduced order models (NM-ROMs). In NM-ROMs, nonlinear manifolds are found by training sparse and shallow autoencoders, which can be costly for large-scale problems. To alleviate the training cost of the sparse and shallow autoencoders, the authors decomposed the domain into subdomains, resulting in the reduction of the number of learnable parameters of autoencoders for each subdomain and the size of the training data. In \cite{romor2023non}, the authors introduced convolutional decoders and compressed decoders. The convolutional decoder learns a latent representation of the dynamical system. There is no constraint on the structure of the convolutional decoder, giving us the opportunity to test various deep learning models. The compressed decoder learns mapping from latent space to subsampled data. The latent space coordinates are found by minimizing subsampled residuals, and then the latent space coordinates are taken as input for the convolutional decoder to predict the solution. In this case, subsampled residuals are only used to find latent space coordinates, and the convolutional decoder computes full-order outputs.

In contrast, non-intrusive methods do not involve solving governing equations. Dynamic identification (DI) methods presented in \cite{fries2022lasdi, he2023glasdi, bonneville2024gplasdi} identify dynamics in latent space using regression method in a least--squares sense. The Gappy proper orthogonal decomposition (POD) method developed by \cite{everson1995karhunen} approximates a solution with a linear subspace spanned by the POD basis matrix. The gappy POD method finds generalized coordinates, which are POD basis coefficients that minimize the error between the predicted and gappy data. This method was successfully applied to unsteady flow reconstruction \cite{willcox2006unsteady}. Linear subspace-based approaches can represent data with fast decaying Kolmogorov N-width, such as elliptic and parabolic equations. However, such methods are not effective in representing data with slowly decaying Kolmogorov N-width, such as linear transport problems \cite{ohlberger2015reduced} and wave equation \cite{greif2019decay}. For digital twins of problems with Kolmogorov N-width decaying slowly, e.g., construction of earthquake digital twin model for source characterization, we need more general method. Since seismic waves are solutions to the wave equation, conventional Gappy POD does not perform well in terms of data reconstruction. The dynamic mode decomposition (DMD) method is a data-driven and non-intrusive method developed by Peter Schmid \cite{schmid2010dynamic, kutz2016dynamic} that approximates local dynamic systems from the dynamic modes found using eigenvectors and eigenvalues of the reduced system obtained through singular value decomposition (SVD) of the dynamic system.

 An alternative method is to determine approximate solutions via machine learning. Physics-informed neural networks (PINNs) \cite{cai2021physics_fluid, cai2021physics_thermal, cuomo2022scientific, raissi2019physics} are trained by minimizing loss between output data and training data together with residuals of given governing equations. PINNs predict solutions directly based on temporal and spatial variables and can be adapted to restore data when sparse measurement data is given to address the challenge of reconstructing gappy data. Specifically, these neural network models use time and spatial parameters in conjunction with the observed data to generate the requisite solutions.

There is also an auto-encoder-based approach to reconstruct data. When gappy data is given as an input, a model outputs reconstructed data directly. In \cite{gundersen2021semi}, the authors presented a conditional variational auto-encoder (CVAE)-based method by conditioning on measurements, simultaneously predicting the solution and uncertainty. Their model reduces the input data to latent space variables, which describe probability distributions. The probability distribution in the latent space is found with the measurement data. Then, solutions can be produced from observations and latent space variables. In \cite{nair2020leveraging}, a deep neural network is used to predict the reduced state of full-order data from sensor measurements directly. Then, the full-order data is reconstructed by multiplying POD basis matrix by the reduced state. 
In \cite{gundersen2021semi} and \cite{nair2020leveraging}, measurement data is an input argument of the neural networks and is used during the forward pass during inference time. However, our proposed method does not take measurement data as the input of the neural network model. In our method, we solve the minimization problem between the measurement data and the model prediction to find generalized coordinates in a latent space.

We developed a novel gappy data reconstruction method called Gappy auto-encoder (AE) to represent physics field data for digital twin purposes. The proposed method learns a nonlinear mapping between the latent and original data spaces using an auto-encoder. Then, we find generalized coordinates in the latent space by minimizing the error in $l_2$ norm between gappy measured data and the decoder output. It is categorized into a data-driven and non-intrusive method. Compared to the Gappy POD method, our Gappy AE method represents data using a nonlinear manifold, overcoming the limitations of linear subspace-based methods. Replacing a linear subspace basis matrix with an auto-encoder requires special care, which is not trivial, such as determining sparse sampling locations, having sparse and shallow decoder structure, and creating the subnet that only contains active nodes and weights as detailed in Section 3 and 4 of \cite{kim2022fast}. The purpose of these non-trivial things is to compute the Jacobian of the decoder quickly which is used in Gauss-Newton iterations for finding generalized coordinates in a latent space. The major contributions of our method are summarized as follows:

\begin{enumerate}
    \item Gappy AE demonstrates superior solution approximations compared to Gappy POD, even with a reduced latent space dimension.
    \item Gappy AE remains effective even in scenarios where measurements are extremely sparse.
    \item Gappy AE framework employs four distinct sampling algorithms, and their respective accuracy is assessed and compared.
\end{enumerate}

The paper is structured as follows: Section \ref{sec:methods} provides an overview of the background materials and our proposed method. We delve into the traditional linear subspace-based approach, the Gappy POD, in Section \ref{sec:gappyPOD}. The Gappy AE method is the focus of Section \ref{sec:gappyAE}, which encompasses three key aspects: Section \ref{sec:NM} explores the concept of a nonlinear manifold-based data representation, Section \ref{sec:NN} delves into the structure of the auto-encoder, and Section \ref{sec:gappy} explains the Gappy AE algorithm. In Section \ref{sec:sampling}, we describe sampling on the residual term and elaborate on the sampling algorithms for measurement location, followed by a showcase of numerical examples in Section \ref{sec:results}. Finally, Section \ref{sec:discussion-conclusion} concludes the paper.

\nomenclature[01]{$\nspacedof$}{Full order model space dimension}
\nomenclature[02]{$\nbasisspace$}{Reduced space dimension}
\nomenclature[03]{$\nbasisres$}{The number of samples}
\nomenclature[04]{$\nparam$}{The number of parameters}
\nomenclature[05]{$\ntimedof$}{The number of time steps}

\nomenclature[06]{$\sol\in\RR{\nspacedof}$}{Solution}
\nomenclature[07]{$\solArg{ref}\in\RR{\nspacedof}$}{Reference solution}
\nomenclature[08]{$\redsolapprox\in\RR{\nbasisspace}$}{Generalized coordinates, or reduced solution}

\nomenclature[09]{$\Ubold\in\RR{\ndof\times\nparam(\ntimedof+1)}$}{Left singular matrix}
\nomenclature[10]{$\Vbold\in\RR{\nparam(\ntimedof+1)\times\nparam(\ntimedof+1)}$}{Right singular matrix}
\nomenclature[11]{$\Sigmabold\in\RR{\nparam(\ntimedof+1)\times\nparam(\ntimedof+1)}$}{Singular values matrix}

\nomenclature[12]{$\basismatspace\in\RR{\nspacedof\times\nbasisspace}$}{Basis matrix with $\nbasisspace\ll\nspacedof$}
\nomenclature[13]{$\scaledDecoder: \RR{\nbasisspace} \rightarrow \RR{\nspacedof}$}{Nonlinear function with $\nbasisspace\ll\nspacedof$}

\nomenclature[14]{$\lVert \boldsymbol{T} \rVert_F$}{Frobenius norm of a matrix $\boldsymbol{T}$}
\nomenclature[15]{$\boldsymbol{T}^\dagger$}{Moore--Penrose inverse of a matrix $\boldsymbol{T}$}
\nomenclature[16]{$\mathbf{J}(\redsolapprox)$}{Jacobian of $\scaledDecoder(\redsolapprox)$}
\nomenclature[17]{$\samplemat\in\RR{\nbasisres\times\ndof}$}{Sampling matrix with $\nbasisres \geq \nbasisspace$}

\nomenclature[18]{$\mathbf{r} \in \RR{\nspacedof}$}{Residual}
\nomenclature[19]{$\hat{\mathbf{r}} \in \RR{\nbasisres}$}{Reduced residual}
\nomenclature[20]{$\basismatspace_r \in \RR{\nspacedof\times\nbasisres}$}{Residual POD basis matrix}

\printnomenclature[5cm]

\section{Methods}\label{sec:methods}
\subsection{Gappy POD}\label{sec:gappyPOD}
The Gappy POD method applies linear subspace-based solution representation, with the solution data approximated as
\begin{equation}\label{eq:LSsolution} 
    \sol \approx \solapprox= \solArg{ref} + \basismatspace\redsolapprox, 
\end{equation} 
where $\solArg{ref}\in\RR{\nspacedof}$ denotes a reference solution, $\basismatspace \defeq [\basisvecspace_1 \cdots \basisvecspace_{\nbasisspace}]\in\RR{\nspacedof\times\nbasisspace}$ denotes a basis matrix, and $\redsolapprox\in\RR{\nbasisspace}$  denotes the generalized coordinates.

The proper orthogonal decomposition (POD) method is commonly used to construct a basis matrix, $\basismatspace$. In POD method \cite{berkooz1993proper}, $\basismatspace$ is obtained by truncating a singular value decomposition (SVD) of a snapshot matrix, $\snapshots\defeq\bmat{\solArg{0}^{\param_1} - \solArg{ref} &
\cdots & \solArg{\ntimedof}^{\param_{\nparam}}- \solArg{ref}}\in\RR{\ndof\times\nparam(\ntimedof+1)}$, where $\solArg{n}^{\param_k}$ is a solution state at $n$-th time step with parameter $\param_k$ for $n\innat{\ntimedof}$ and $k\innat{\nparam}$. It pertains to the statistical analysis technique known as principal component analysis \cite{hotelling1933analysis} and the stochastic analysis concept called Karhunen--Lo\`{e}ve expansion \cite{loeve1955}. The POD method computes its thin SVD: 
     \begin{equation}\label{eq:SVD} 
       \snapshots = \Ubold\Sigmabold\Vbold^T,
     \end{equation} 
where $\Ubold\in\RR{\ndof\times\nparam(\ntimedof+1)}$ and $\Vbold\in\RR{\nparam(\ntimedof+1)\times\nparam(\ntimedof+1)}$ are orthogonal matrices and $\Sigmabold\in\RR{\nparam(\ntimedof+1)\times\nparam(\ntimedof+1)} $ is a diagonal matrix with singular values on its diagonals. The leading $\nbasisspace$ columns of $\Ubold$ are selected to set $\basismatspace$ (i.e., $\basismatspace = \bmat{\ubold_1 & \cdots & \ubold_{\nbasisspace}}$, where $\ubold_k$ is $k$-th column vector of $\Ubold$). The basis matrix from POD method, $\basismatspace$ is a solution to the minimization problem among $\Tilde{\basismatspace}\in\RR{\ndof\times\nbasisspace}$ with orthonormal columns
\begin{equation}
    \basismatspace = \argmin{} \|\snapshots - \Tilde{\basismatspace}\Tilde{\basismatspace}^T\snapshots \|_F^2.
\end{equation}
For more detailed information on the POD method, please refer to \cite{hinze2005proper,kunisch2002galerkin}.

\begin{figure}[h]
  \centering
  \includegraphics[width=0.9\textwidth]{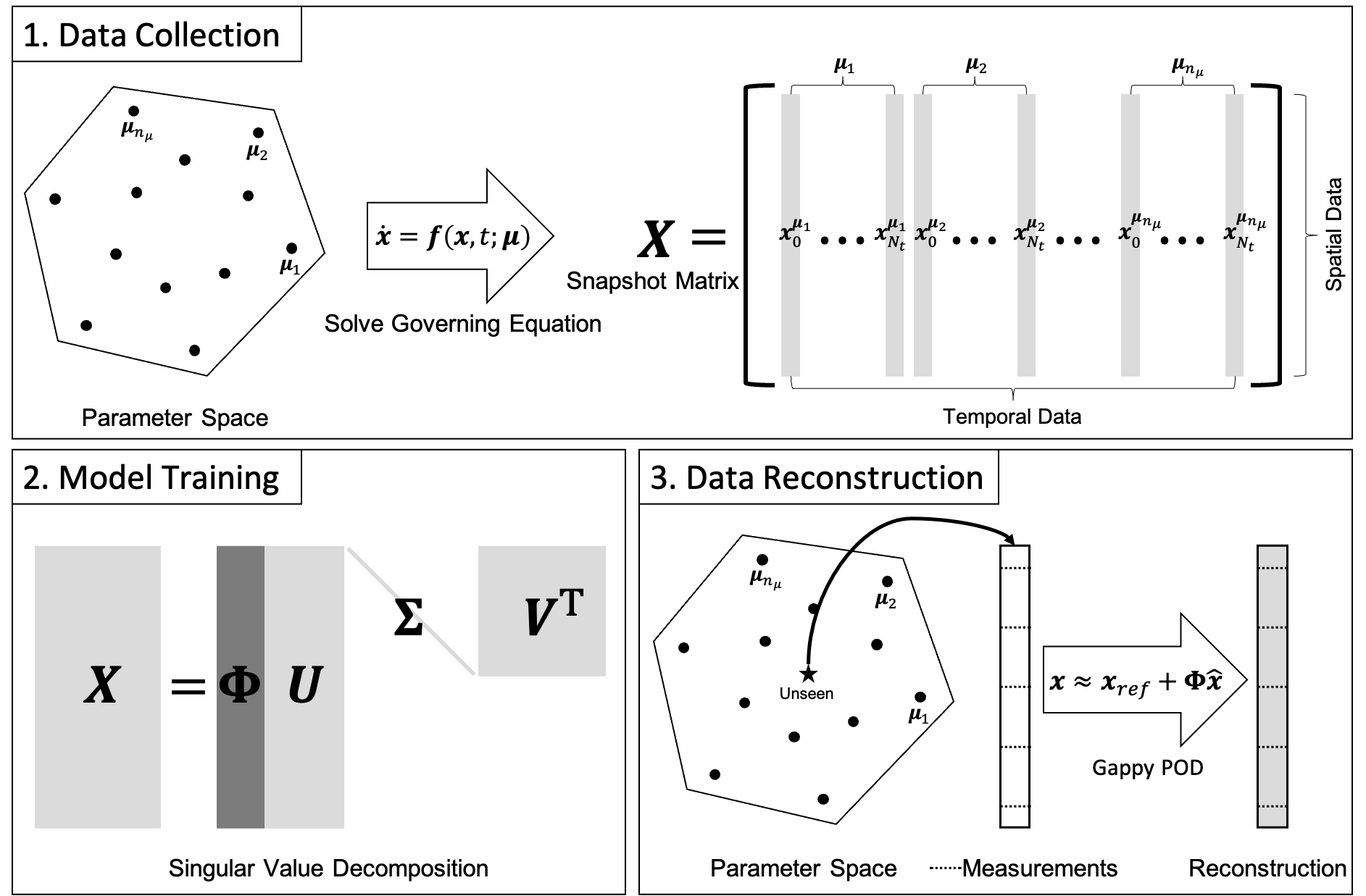}
  \caption{Graphical Summary of Gappy POD process from 1 to 3.}
  \label{fg:gappyPOD}
\end{figure}

For brevity and readability, $n$ and $\param_k$ are omitted in the notation $\solArg{n}^{\param_k}$. We can set a minimization problem between the measurement and solution data approximation given the sparse measurement data. By solving the minimization problem, we find the generalized coordinates, $\redsolapprox$. The Gappy POD first approximates $\sol \in \RR{\nspacedof}$ as
\begin{equation}
    \sol \approx \solapprox = \basismatspace \redsolapprox,
\end{equation}
assuming $\solArg{ref}=\mathbf{0}$. Solving the following minimization problem
\begin{equation}
    \redsolapprox=\argmin{\reddummy\in\RR{\nbasisspace}}\lVert \samplemat (\sol-\basismatspace \reddummy) \rVert_2^2
\end{equation}
with given sampling matrix, $\samplemat\in\RR{\nbasisres\times\ndof}$, that extracts the rows of matrix and vector corresponding to measurement points, where $\nbasisres$ is the number of samples and $\nbasisres \geq \nbasisspace$ gives us
\begin{equation}
    \redsolapprox = (\samplemat\basismatspace)^{\dagger}\samplemat \sol,
\end{equation}
and then we have
\begin{equation}\label{eq:gappyPODcomputation}
    \sol  \approx \basismatspace \redsolapprox = \basismatspace (\samplemat\basismatspace)^{\dagger}\samplemat \sol,
\end{equation}
where 
$\samplemat \sol$ is replaced with the sparse measurement data. 
The graphical summary of Gappy POD process is shown in Fig. \ref{fg:gappyPOD}.
 \clearpage

\subsection{Gappy Auto-Encoder}\label{sec:gappyAE}
\begin{figure}[h]
  \centering
  \includegraphics[width=0.9\textwidth]{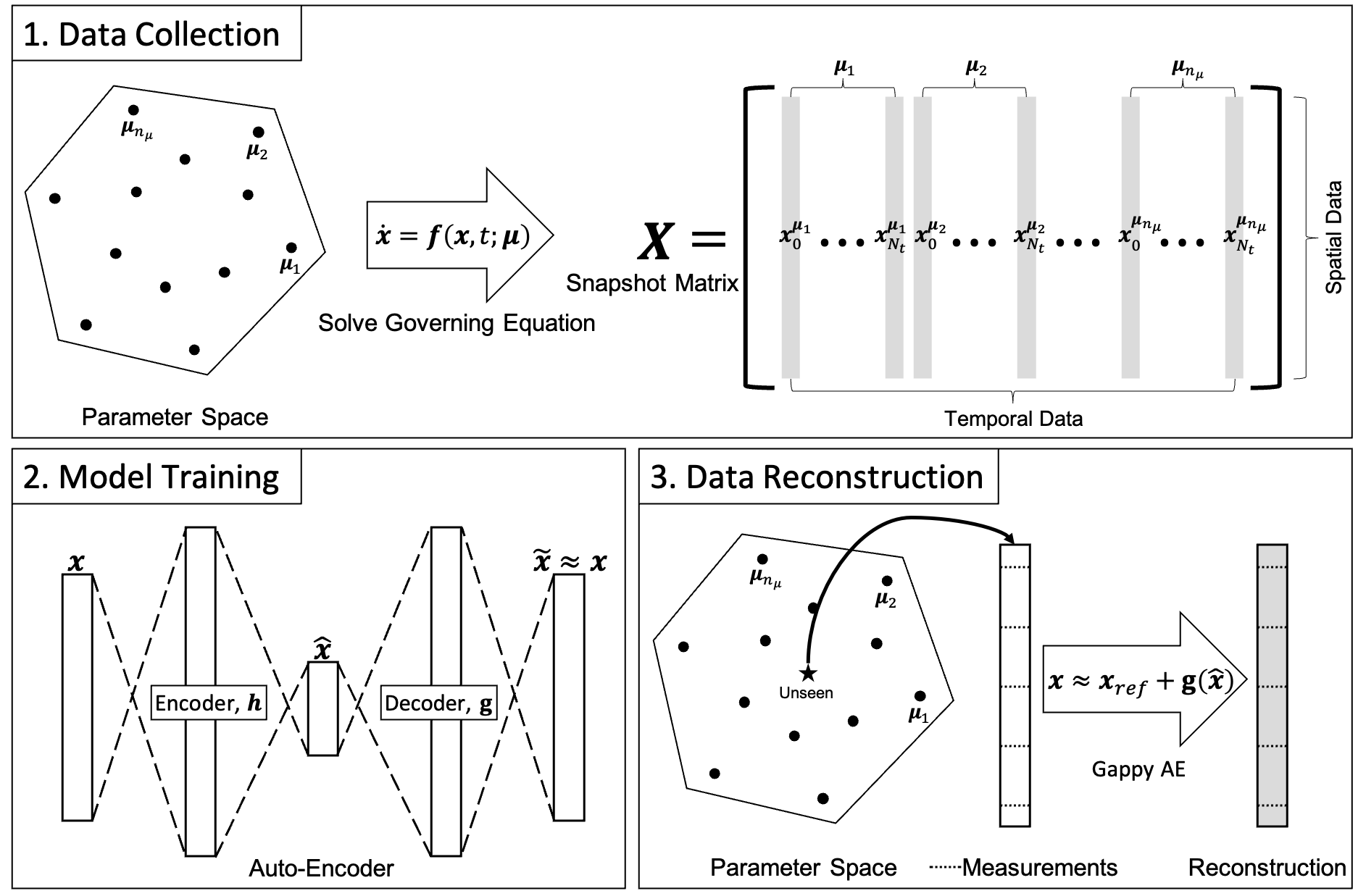}
  \caption{Graphical Summary of Gappy AE process from 1 to 3.} 
  \label{fg:gappyAE}
\end{figure}

Gappy AE uses nonlinear manifold solution representation described in Section \ref{sec:NM}. The nonlinear manifold is found by training the auto-encoder in an unsupervised fashion as presented in \ref{sec:NN}. In Section \ref{sec:gappy}, gappy procedures for finding points in the reduced space with given measurements are given. A graphical summary of the Gappy AE process is shown in Fig. \ref{fg:gappyAE}. 
 \clearpage

\subsubsection{Nonlinear manifold solution representation}\label{sec:NM}
The Gappy AE utilizes a nonlinear manifold $\spatialSubspace \defeq \{\scaledDecoder\left(\reddummy\right)|\reddummy \in \RR{\nbasisspace}\}$, where $\scaledDecoder: \RR{\nbasisspace} \rightarrow \RR{\nspacedof}$ with $\nbasisspace\ll\nspacedof$ denoting a nonlinear function that maps a latent space of dimension $\nbasisspace$ to the full order model space of dimension $\nspacedof$. In other words, the Gappy AE estimates the solution by approximating it within a trial manifold as
\begin{equation}\label{eq:spatialNMROMsolution} 
  \sol \approx \solapprox= \solArg{ref} + \scaledDecoder \left(\redsolapprox
  \right), 
\end{equation} 
where $\redsolapprox\in\RR{\nbasisspace}$ denotes the generalized coordinates. The initial condition for the generalized coordinate, $\redsolapproxArg{0}\in\RR{\nbasisspace}$, is given by $\redsolapproxArg{0}=\scaledEncoder\left(\solArg{0}-\solArg{ref}\right)$, where $\scaledEncoder \approx \scaledDecoder^{-1}$ (i.e., $\sol-\solArg{ref} \approx \scaledDecoder\left(\scaledEncoder\left(\sol-\solArg{ref}\right)\right)$). The details of the nonlinear functions, $\scaledEncoder$ and $\scaledDecoder$, are presented in Section \ref{sec:NN}. 

\subsubsection{Sparse Shallow Auto-Encoder}\label{sec:NN}
Auto-Encoder is a feed-forward neural network model that learns identity mapping. The auto-encoder consists of an encoder, $\scaledEncoder:\RR{\nspacedof} \rightarrow \RR{\nbasisspace}$, and a decoder, $\scaledDecoder: \RR{\nbasisspace} \rightarrow \RR{\nspacedof}$. The nonlinear activation functions in the auto-encoder make them nonlinear. The encoder maps high dimensional data, $\sol \in \RR{\nspacedof}$ to low dimensional data, $\redsolapprox \in \RR{\nbasisspace}$ and the decoder maps back the low dimensional data to the high dimensional data. The decoder is used for nonlinear mapping, $\scaledDecoder(\redsolapprox)$, mentioned in Section \ref{sec:NM}.

The general structure of the encoder and decoder can vary, but for our study, we use a sparse shallow network as outlined in Section 3.2 of \cite{kim2022fast}. The sparsity in the network enhances efficiency by eliminating connections in the layers. However, when constructing neural networks for physical data with higher dimensions, caution must be exercised to ensure that the spatial data organization aligns with the sparse connection. In other words, spatially near nodes share part of the connections, whereas far-away nodes in the space domain do not share the connections even if two nodes are neighbors. In this work, the encoder and decoder have three layers. 
The encoder layers are fully connected and a nonlinear activation function is applied in the hidden layer, while the output layer lacks a nonlinear activation function. For the decoder architecture, the nodes between the input and hidden layer are fully connected with a nonlinear activation function, however, the nodes between the hidden and output layer are sparsely connected without an activation function. We have introduced two hyper-parameters, $b$, and $\delta b$, to design a sparsity structure. $b$ denotes the size of a block to compute for one output node and $\delta b$ defines the amount of shift that the block moves. In our work, a swish activation function is used.
For more information on the neural network architecture, please refer to \cite{kim2022fast}. 

A sparse and shallow decoder can efficiently compute a subset of the output because the active nodes in a hidden layer constitute a subset of the entire set of hidden nodes. Given that we exclusively analyze the sparse measurement data, we only need to consider the subset of output nodes generated by the decoder. This approach enables leveraging the inherent sparsity of measurements. In essence, the sparse and shallow neural network structure can be streamlined into a more compact network by retaining only the active nodes and edges, allowing computing of the output nodes associated with measurement points.

\subsubsection{Gappy Procedures}\label{sec:gappy}
Minimizing error between measurements and decoder output leads to a nonlinear problem, which can be solved by the Gauss-Newton method to find latent space coordinates. We first define $\mathbf{n}\in \RR{\nspacedof}$ as
\begin{equation}
    n_j=\begin{cases}
        1, & \text{if $x_j$ is known}\\
        0, & \text{if $x_j$ is missing}
        \end{cases}
\end{equation}
where $\sol \in \RR{\nspacedof}$ is a vector that represents the solution at a given time and parameter and $x_j$ denotes $j$-th component of $\sol$. Then, a diagonal matrix, $\mathbf{N}\in\RR{\nspacedof\times\nspacedof}$ is constructed with elements of $\mathbf{n}$ vector on the diagonal and removing zero columns gives us $\mathbf{Z} \in \RR{\nspacedof\times\nbasisres}$, where $\nbasisres$ is the number of known values. Now, we define an error, $\epsilon \in \RR{}$ as 
\begin{equation}
    \epsilon = \lVert \samplemat(\sol-\solapprox) \rVert_2^2,
\end{equation}
where $\samplemat\sol$ is a measurement data and $\solapprox$ is prediction given by
\begin{equation}
    \solapprox=\sol_{ref}+\scaledDecoder(\mathbf{b}),
\end{equation}
where 
$\mathbf{b} \in \RR{\nbasisspace}$ is a generalized coordinate to be found.\par

Minimizing $\epsilon$ gives us a generalized coordinate $\mathbf{b}$ through the Gauss--Newton method. At the $i$-th iteration, first, we linearize $\samplemat(\sol-\solapprox)$ at $\mathbf{b}_i$ as
\begin{equation}
    \samplemat(\sol-\sol_{ref}-\scaledDecoder(\mathbf{b}_i)-\mathbf{J}\Delta\mathbf{b}_i),
\end{equation}
where $\mathbf{J}=\frac{\partial \solapprox}{\partial\mathbf{b}}$ and $\Delta\mathbf{b}_i=\mathbf{b}_{i+1}-\mathbf{b}_i$. Then, the linearized error $\epsilon_{lin}$ becomes
\begin{equation}
    \epsilon_{lin} =(\sol-\sol_{ref}-\scaledDecoder(\mathbf{b}_i)-\mathbf{J}\Delta\mathbf{b}_i)^T\samplematNT\samplemat(\sol-\sol_{ref}-\scaledDecoder(\mathbf{b}_i)-\mathbf{J}\Delta\mathbf{b}_i).
\end{equation}
Imposing $\frac{\partial \epsilon_{lin}}{\partial \mathbf{b}}=0$, which is given by
\begin{align}
    \frac{\partial \epsilon_{lin}}{\partial \mathbf{b}} &= 2\mathbf{J}^T\samplematNT\samplemat(\sol-\sol_{ref}-\scaledDecoder(\mathbf{b}_i)-\mathbf{J}\Delta\mathbf{b}_i) \\
    &= 0
\end{align}
gives us update $\Delta\mathbf{b}_i$
\begin{align}
    \Delta\mathbf{b}_i &= (\mathbf{J}^T\samplematNT\samplemat\mathbf{J})^{-1}\mathbf{J}^T\samplematNT\samplemat(\sol-\sol_{ref}-\scaledDecoder(\mathbf{b}_i))\\
    &= (\samplemat\mathbf{J})^{\dagger}\samplemat(\sol-\sol_{ref}-\scaledDecoder(\mathbf{b}_i)).
\end{align}
Thus, we have 
\begin{equation}
    \mathbf{b}_{i+1}=\mathbf{b}_i+ (\samplemat\mathbf{J})^{\dagger}\samplemat(\sol-\sol_{ref}-\scaledDecoder(\mathbf{b}_i)).
\end{equation}
The above steps are iterated until the error is minimized. Let us write $\redsolapprox$ as the generalized coordinate at the final iteration. Then $\solapprox$ is computed using $\solapprox=\sol_{ref}+\scaledDecoder(\redsolapprox)$ to complete the reconstruction.

\subsection{Sampling Algorithms}\label{sec:sampling}
To reconstruct incomplete data using sparse measurements, we optimize the sensor locations to minimize reconstruction errors. We conducted experiments with the Gappy AE and POD algorithms using various sampling methods, namely uniform sampling, Latin hypercube sampling (LHS) \cite{mckay2000comparison}, discrete empirical interpolation method (DEIM) \cite{chaturantabut2010nonlinear, drmac2016new, drmac2018discrete, carlberg2013gnat, choi2020sns}, and S-OPT \cite{shin2016nonadaptive,lauzon2022s}.

In practical applications, sensor placement options are often constrained by factors such as product design, operational considerations, and maintenance requirements. We assume the flexibility to position sensors anywhere for our sampling algorithm tests.

For the Gappy POD method, we use the POD basis matrix for DEIM and S-OPT sampling algorithms; however, the Gappy AE does not have such POD basis matrix. To apply DEIM and S-OPT sampling algorithms to the Gappy AE, we introduce a residual, $\mathbf{r} \in \RR{\nspacedof}$, defined by
\begin{equation}
    \mathbf{r}=\sol - \scaledDecoder(\redsolapprox),
\end{equation}
assuming $\sol_{ref}=0$. Following the GNAT-SNS approach introduced in \cite{choi2020sns}, the residual is approximated by
\begin{equation}\label{eq:ResApprox}
    \mathbf{r}\approx\Tilde{\mathbf{r}}=\basismatspace_r\hat{\mathbf{r}},
\end{equation}
where $\basismatspace_r \in \RR{\nspacedof\times\nbasisres}$ is POD basis matrix from the residual snapshot matrix and $\hat{\mathbf{r}} \in \RR{\nbasisres}$ is a reduced residual. The residual snapshot is constructed by concatenating residual vectors during iterations when minimizing $\mathbf{r}$ for the training data. Then, applying SVD to the residual snapshot and choosing leading $\nbasisres$ left singular vectors yield the residual basis matrix, $\basismatspace_r$.

Next, we define $\samplemat\defeq[\unitvecArg{p_1},\ldots,\unitvecArg{p_{\nbasisres}}]^T \in\RR{\nbasisres\times\ndof}$, $\nbasisspace \leq \nbasisres \ll \ndof$, is the sampling matrix and $\unitvecArg{p_i}$ is  the $p_i$th column of the identity matrix  $\identity{\ndof}\in\RR{\ndof\times\ndof}$. $\samplemat$ extracts rows corresponding to sampling points from matrices or vectors. Then, solving the following equation
\begin{align}
    \samplemat \mathbf{r} &= \samplemat \mathbf{\Tilde{r}} \\
                          &= \samplemat \basismatspace_r\hat{\mathbf{r}},
\end{align}
we obtain
 \begin{equation}\label{eq:HR-generalizedcoordinates}
   \redres = (\samplemat\basismatres)^{-1}\samplemat\mathbf{r}.
 \end{equation}
Therefore, we have
\begin{equation}
    \mathbf{r}  \approx \basismatres (\samplemat\basismatres)^{-1}\samplemat \mathbf{r}.
\end{equation}
Then, the residual minimization problem given by
\begin{equation}
    \redsolapprox=\argmin{\reddummy\in\RR{\nbasisspace}}\lVert \basismatres (\samplemat\basismatres)^{-1}\samplemat (\sol - \scaledDecoder(\reddummy)) \rVert_2^2
\end{equation}
solves $\redsolapprox$ using the Gauss--Newton method. The update for each iteration is given by
\begin{equation}
    \redsolapprox_i=\redsolapprox_{i-1}+\{(\samplemat\basismatres)^{-1}\samplemat\mathbf{J}(\redsolapprox_{i-1})\}^{\dagger}(\samplemat\basismatres)^{-1}\samplemat(\sol-\scaledDecoder(\redsolapprox_{i-1})),
\end{equation}
where $\mathbf{J}(\redsolapprox)$ is a Jacobian of $\scaledDecoder(\redsolapprox)$.

\subsubsection{Discrete Empirical Interpolation Method (DEIM)}\label{sec:DEIM}
The DEIM algorithm employed here is the so-called oversampled DEIM. The oversampled DEIM differs from the original DEIM approach and aligns more closely with the sampling technique presented in \cite{carlberg2011efficient}. Moreover, unlike the original DEIM, the oversampled DEIM allows a larger number of sample points to be selected than the number of POD basis.

Given POD basis matrix, $\mathbf{\Phi}=[\boldsymbol{\phi}_1,\cdots,\boldsymbol{\phi}_p] \in \RR{N\times p}$, the algorithm selects indices such that the reconstruction error bound is minimized in a greedy sense. The first sampling index is chosen at the largest absolute value of $\boldsymbol{\phi}_1$. Next, the algorithm proceeds to iterate through the remaining columns of the $\boldsymbol{\Phi}$ for the $j$th iteration with $j>1$, approximating
\begin{equation}
\Tilde{\boldsymbol{\phi}}_j=\basismatspace_{1:j-1}(\samplemat_{1:j-1}\basismatspace_{1:j-1})^{\dagger}\samplemat_{1:j-1}\boldsymbol{\phi}_j,
\end{equation}
where $\basismatspace_{1:j-1}=[\boldsymbol{\phi}_1,\cdots,\boldsymbol{\phi}_{j-1}]$ and $\samplemat_{1:j-1}$ is a sampling matrix that extracts selected rows (i.e., selected indices) up to $(j-1)$th iteration. Then, the next index is selected, at which the absolute value of the error
\begin{equation}
    \boldsymbol{\epsilon}_j=\boldsymbol{\phi}_j - \Tilde{\boldsymbol{\phi}}_j
\end{equation}
is the largest. When oversampling, we iterate the above process multiple times as we update sampling indices for a given $\boldsymbol{\phi}_j$. Implementation-wise, we follow Algorithm 3.1 of \cite{lauzon2022s} for the oversampled DEIM.

\subsubsection{S-OPT}\label{sec:SOPT}
We introduce the S-OPT sampling algorithm presented in \cite{shin2016nonadaptive,lauzon2022s}. As in Section \ref{sec:DEIM}, we denote $\basismatspace$ as the POD basis matrix, $\samplemat$ as the sampling matrix, and $\sol$ as the ground truth data to be reconstructed. According to Theorem 3.1 of \cite{lauzon2022s}, we can write the reconstruction error as 
\begin{equation}
    \|(\boldsymbol{I}-\basismatspace(\samplemat\basismatspace)^{\dagger}\samplemat)\sol\|_2^2=\|(\boldsymbol{I}-\basismatspace\basismatspace^T)\sol\|_2^2+\|\boldsymbol{\epsilon}\|_2^2,
\end{equation}
where $\boldsymbol{\epsilon}=((\samplemat\basismatspace)^T\samplemat\basismatspace)^{-1}(\samplemat\basismatspace)^T\samplemat(\boldsymbol{I}-\basismatspace\basismatspace^T)\sol$. Since we do not know the ground truth of $\sol$ in practice, the true optimum $\samplematNT^{\ast}$ cannot be found. We use training data to find $\samplematNT$ and this is quasi-optimal. The S-OPT algorithm finds sampling points to have the smallest $\boldsymbol{\epsilon}$ under the condition of given training data of $\sol$, corresponding to maximizing the column orthogonality of $\samplemat\basismatspace$ and determinant of $(\samplemat\basismatspace)^T\samplemat\basismatspace$. In \cite{shin2016nonadaptive}, the quantity $S:\RR{N \times p} \rightarrow [0,1]$ is defined as
\begin{equation}
    S(\boldsymbol{A})\coloneqq \frac{\sqrt{\det{\boldsymbol{A}^T\boldsymbol{A}}}}{\Pi_{i=1}^p\|\boldsymbol{\alpha}_i\|},
\end{equation}
where $\boldsymbol{A}$ is a $N \times p$ matrix and  $\boldsymbol{\alpha}_i$ is the $i$th column of $\boldsymbol{A}$ assuming $\|\boldsymbol{\alpha}_i\|\neq 0$. Maximizing $S(\boldsymbol{A})$ provides maximal orthogonality of $\boldsymbol{A}$ and determinant of $\boldsymbol{A}^T\boldsymbol{A}$. 
In the context of sampling algorithms in the Gappy POD and AE, the S-OPT seeks the optimal sampling matrix $\boldsymbol{Z}$ that maximizes $S$, i.e., 
\begin{equation}
\boldsymbol{Z}_{\text{S-OPT}} = \underset{\boldsymbol{Z}}{\text{argmin}} \hspace{4pt}S(\samplemat \basismatspace) 
\end{equation}
The S-OPT sampling algorithm is presented in Section 3 of \cite{shin2016nonadaptive}. The source code of the S-OPT algorithm is available at \cite{doecode_24508}.

\section{Numerical Results}\label{sec:results}
We used three normalized example problems, diffusion, radial advection, and wave problems, for testing the Gappy AE algorithm. The Gappy AE results are shown in this section. Due to page constraints, the Gappy POD results are presented in Appendix B of \cite{appendix}. The Python codes that are used to produce the results with noiseless measurement assumption are available at \url{https://github.com/youngkyu-kim/GappyAE}. The codes for noisy measurement cases are omitted because the only difference is the line for adding noise to the measurement data.
Open source finite element method (FEM) solver, called  MFEM \cite{mfem,mfem-web} is used to generate simulation data. The auto-encoder is trained using the simulation data, where training parameters for each problem are given by
\begin{equation}
    \mu \in \paramDomain_{train}=\{0.8,0.85,0.9,0.95,1,1.05,1.1,1.15,1.2\}
\end{equation} 
and test parameters are given by
\begin{equation}
    \mu \in \paramDomain_{test}=\{\mu_i|\mu_i=0.75+i/100, i=0,1,\cdots,50\}
\end{equation} 
In our numerical examples, the width of the encoder is twice the input size and the latent space dimension varies from 3 to 6. The block size and the amount of shifts are set as 100 and $20$, respectively. The batch size is 240 and the data set is split into $80\%$ for training and $20\%$ for testing. The ADAM optimizer, a variant of stochastic gradient descent (SGD), is used to update the learnable parameters. The initial learning rate is set as 0.001 and reduced by a factor of 10 if learning shows no improvement for 50 epochs. The number of epochs for training is 10000, but if training does not yield improvement after 200 times, the training is stopped. The autoencoders are trained on a single NVIDIA A100 GPU with 40GB of memory. The average training time per epoch is $0.539$ seconds with a latent space dimension set to $6$. The early stopping scheme introduces variability in the training time; however, we enforce a maximum of $10000$ epochs. Consequently, the training time is capped at approximately $90$ minutes. The mean squared error (MSE) loss of the auto-encoder is shown in Fig. 1 of Appendix A in \cite{appendix}. 
For all cases, both training curves and test curves consistently continue to improve and we don’t see the test curves plateau or worsen, which suggests the models are not overfitting. We conclude that the auto-encoder models are not over-fitted to the training data.

The auto-encoder training is repeated 10 times to account for randomness due to random initialization and mini-batch sampling. The statistics of MSE losses are presented in Fig. 2 of Appendix A in \cite{appendix}. The MSE losses decrease as latent space dimensions increase because of more learnable hyper-parameters.

For each numerical problem, we present two cases: measurements with and without noise. We assume white noise and create noisy measurements by adding random numbers with uniform distribution within the range $[-0.01, 0.01)$, which is $\pm 1\%$ of the maximum data value in each example. In our numerical examples, projection error is calculated when all data points are measured with and without noise and used as a low bound of the Gappy AE and POD data reconstruction accuracy. The Gappy AE projection errors are shown in Figs. 3, 4, and 5 of Appendix A in \cite{appendix}. The Gappy POD projection errors are presented in Figs. 26, 27, and 28 of Appendix B in \cite{appendix}. Gappy AE is $\mathcal{O}(10)$ to $\mathcal{O}(100)$ times more accurate than the Gappy POD in terms of projection error. The projection errors without noise are slightly lower than those with noise; the effects of noise in computing projection errors are negligible.

For our numerical examples, the size of the reduced dimension (i.e., latent space dimension for nonlinear manifold or the reduced basis dimension for linear subspace) should be at most half of the number of sample points to avoid numerical round-off errors. If the reduced dimension size is close to the number of sample points, the condition number of the matrix in solving $\redsolapprox$ increases, leading to errors in the calculation of the pseudo-inverse. In addition, the reduced dimension size is greater than that of the intrinsic dimension. Through the numerical examples, the intrinsic dimension is 2, with time $t \in \RRplus{}$ and parameter $\mu \in \RR{}$; the number of sample points is 12. Thus, we limit the dimension size to $\{3,4,5,6\}$.

The computational cost of the online phase of gappy data reconstruction is measured in terms of wall-clock time. Calculations are performed on one CPU of an AMD EPYC 7763 @ 2.45 GHz and DDR4 Memory @ 3200 MT/s. With given $12$ sample points and reduced dimension (i.e., the POD basis or latent space dimensions) of $[3,4,5,6]$, the online phase of the Gappy POD method for each reconstruction step in our example problems takes around $0.1$ ms. However, it takes around $3.0$ to $3.8$ ms for the Gappy AE method to reconstruct the data during the online phase. To compare the online cost of Gappy AE, Gappy POD, and FE simulation, wall-clock time for each time step is measured $10$ times and the mean value is used in Table \ref{tb:online_cost}. The performances of Gappy AE and POD methods in terms of reconstruction error are compared in the following sections.
\begin{table}[h]
\caption{Comparison of online cost of Gappy AE and Gappy POD under noiseless measurement conditions. Note that FE simulation does not reconstruct data but solves governing equations. For data reconstruction, the best-performing cases in terms of accuracy from each numerical example are selected, and the online costs of Gappy AE, Gappy POD, and FE simulation are computed by averaging the mean wall-clock times for each time step over FOM parameters.}\label{tb:online_cost}
\centering
\begin{tabular}{|c|c|ccc|}
\hline
\multirow{2}{*}{Problem Type}     & \multirow{2}{*}{Sensor Placement} & \multicolumn{3}{c|}{Time (ms)}                                                          \\ \cline{3-5} 
                                  &                                   & \multicolumn{1}{c|}{Gappy AE} & \multicolumn{1}{c|}{Gappy POD} & FE simulation          \\ \hline
\multirow{2}{*}{Diffusion}        & Boundary                          & \multicolumn{1}{c|}{3.794}    & \multicolumn{1}{c|}{0.119}     & \multirow{2}{*}{44.481} \\ \cline{2-4} 
                                  & Domain                            & \multicolumn{1}{c|}{3.045}    & \multicolumn{1}{c|}{0.118}     &                         \\ \hline
\multirow{1}{*}{Radial Advection} & Domain                            & \multicolumn{1}{c|}{3.161}    & \multicolumn{1}{c|}{0.121}     & 32.220                   \\ \hline 
\multirow{2}{*}{Wave}             & Boundary                          & \multicolumn{1}{c|}{3.557}    & \multicolumn{1}{c|}{0.121}     & \multirow{2}{*}{59.400}  \\ \cline{2-4} 
                                  & Domain                            & \multicolumn{1}{c|}{3.355}    & \multicolumn{1}{c|}{0.121}     &                          \\ \hline
\end{tabular}
\end{table}

\subsection{Diffusion Problem}\label{sec:diffusion}
\begin{figure}[h]
  \centering
  \subfigure[$\mu_{min}=0.75$]{
  \includegraphics[width=0.9\textwidth]{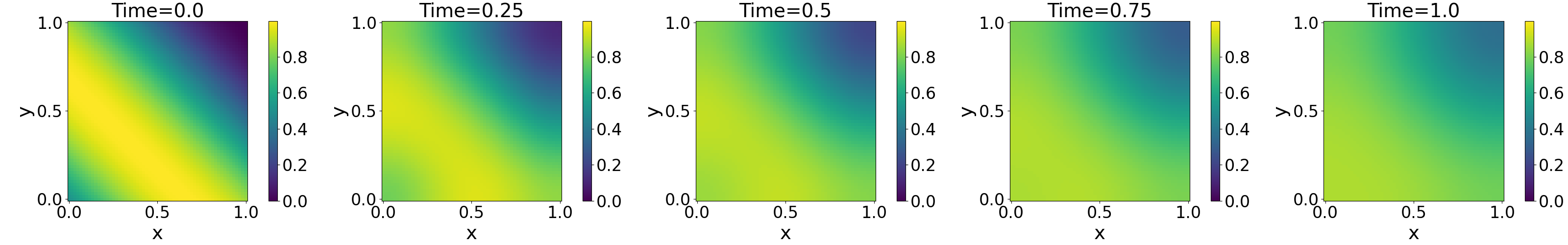}}
  \subfigure[$\mu_{max}=1.25$]{
  \includegraphics[width=0.9\textwidth]{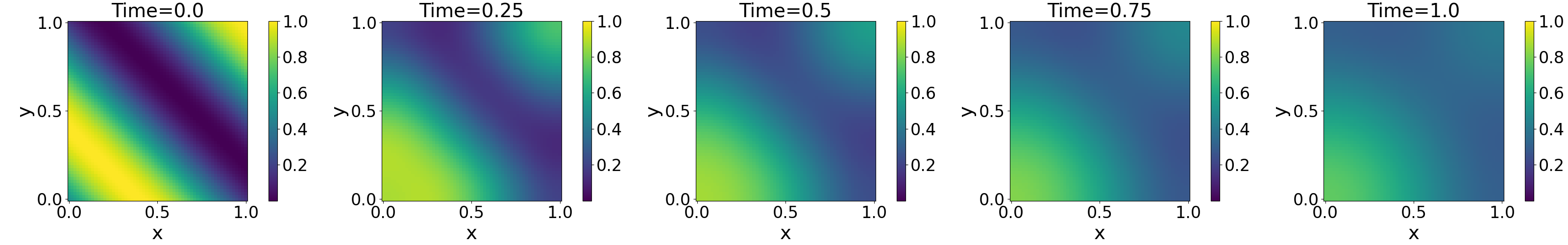}}
  \caption{Diffusion simulation solutions from the initial to the final time for the two endpoints of $\mu$}
  \label{fg:ex16SolTwoParam}
\end{figure}

A 2D parameterized nonlinear diffusion problem is considered
\begin{equation}
    \frac{\partial u}{\partial t}=\nabla\cdot(\kappa+\alpha u)\nabla u,
\end{equation}
where spatial and temporal domains are given by $\mathbf{x}\in\Omega=[0,1]\times[0,1]$ and $t\in[0,1]$, respectively. Here, we set $\kappa = 0.05$ and $\alpha=0.01$. The Neumann boundary condition is imposed
\begin{equation}
\frac{\partial u}{\partial \mathbf{x}}\cdot \boldsymbol{n}=0 \quad \text{on} \quad \partial \Omega,
\end{equation}
where $\boldsymbol{n}$ denotes the unit normal vector. The initial condition is imposed
\begin{equation}
    u(0,\mathbf{x};\mu)=0.5\sin(\mu \|\mathbf{x}\|_2) + 0.5,
\end{equation}
where $\mu$ is a parameter. The spatial domain is discretized into a 64 by 64 square mesh. The diagonally implicit Runge--Kutta with $\Delta t=0.002$ is employed as a time integrator. This problem is from Example 16 of MFEM codes. The source code of the diffusion problem is available at \url{https://github.com/mfem/mfem/blob/master/examples/ex16.cpp}. Figure~\ref{fg:ex16SolTwoParam} shows five snapshots for two extreme parameter values. 

In the diffusion problem, two sensor placement scenarios are presented: on the boundary and in the domain.

 \clearpage

\subsubsection{Sensor Placement on Boundary}\label{sec:diffusion_bndry}
This section considers that the sensors are located at the boundary region only. The latent space dimension of the auto-encoder varies from 3 to 6 and three kinds of sampling methods, i.e., uniform sampling, DEIM, and S-OPT sampling methods, are employed.

\begin{figure}[h]
  \centering
  \subfigure[Average Relative Error (\%)]{
  \includegraphics[width=0.9\textwidth]{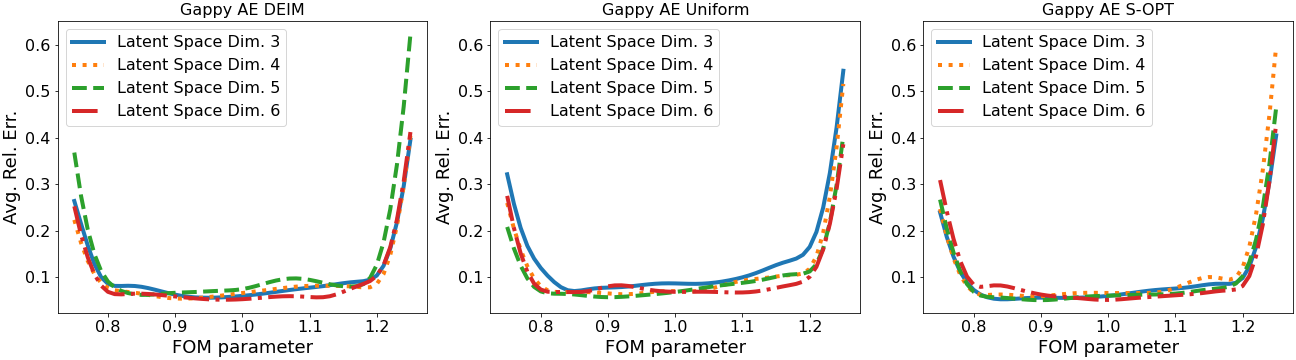}}
  \subfigure[Maximum Relative Error (\%)]{
  \includegraphics[width=0.9\textwidth]{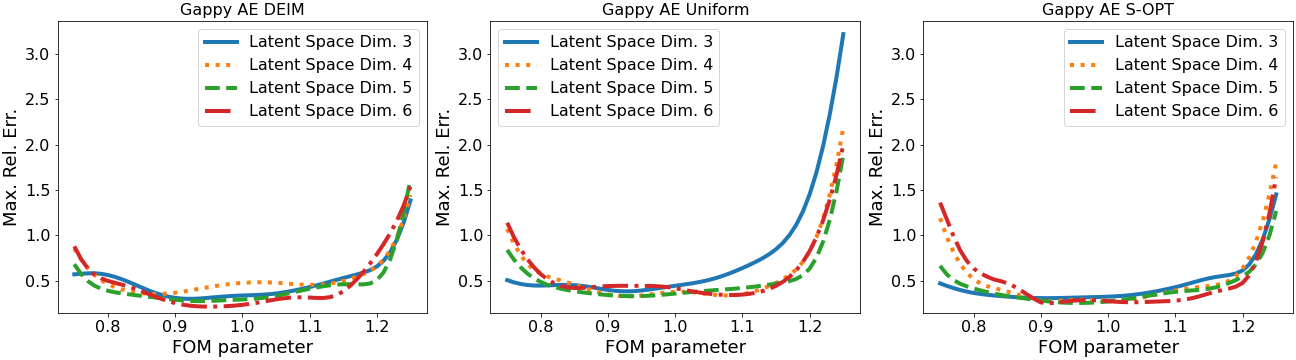}}
  \caption{Accuracy of Gappy AE for diffusion problem with noiseless measurements on the boundary}
  \label{fg:ex16BndryGappyAESize}
\end{figure}

\begin{figure}[h]
  \centering
  \subfigure[Average Relative Error (\%)]{
  \includegraphics[width=0.9\textwidth]{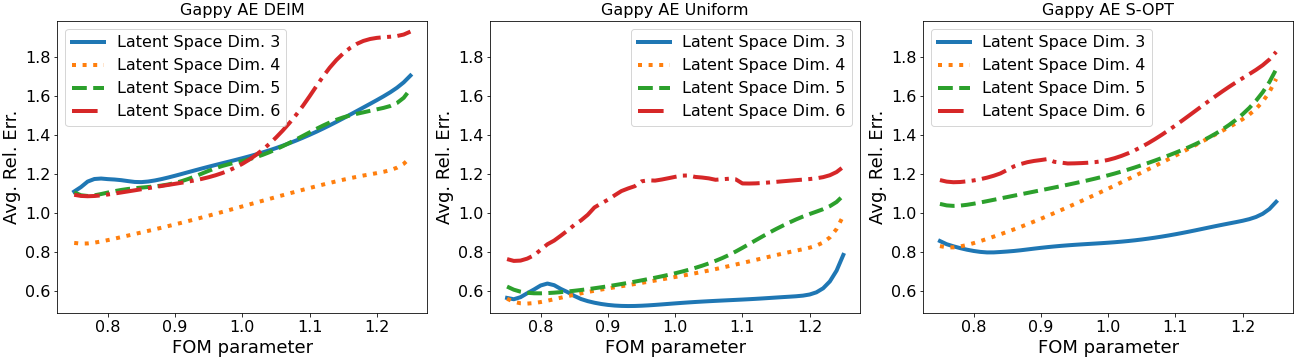}}
  \subfigure[Maximum Relative Error (\%)]{
  \includegraphics[width=0.9\textwidth]{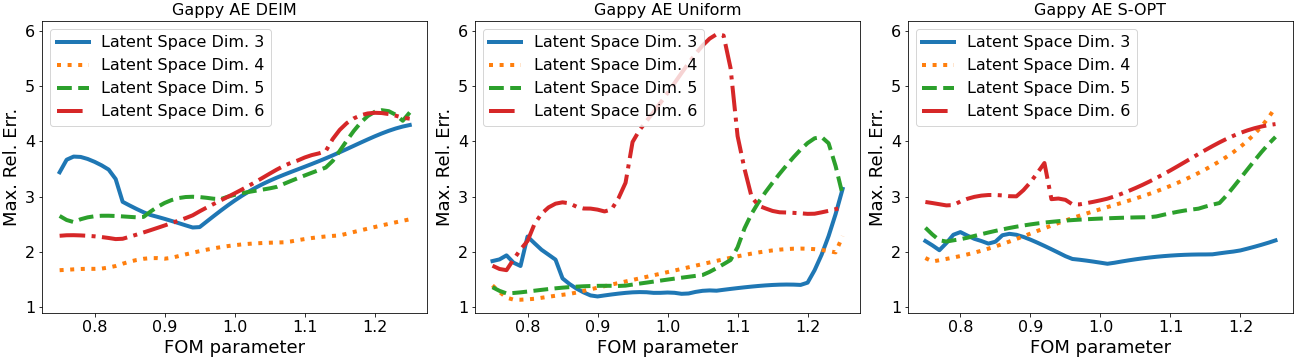}}
  \caption{Accuracy of Gappy AE for diffusion problem with noisy measurements on the boundary}
  \label{fg:ex16BndryNoisyGappyAESize}
\end{figure}

First, we show the effects of the latent space dimensions in Figs. \ref{fg:ex16BndryGappyAESize} and \ref{fg:ex16BndryNoisyGappyAESize}. In Fig. \ref{fg:ex16BndryGappyAESize}, the relative errors are similar over the latent space dimensions. However, with noise in measurement data, as described in Section \ref{sec:results}, the accuracy of Gappy AE is better when the latent space is 3 or 4. We consider that the auto-encoder with a smaller latent space dimension is less sensitive to noise.

 \clearpage

 \clearpage

Next, we choose the best-performing latent space dimension for each sampling method and they are compared in Figs. 6, 7, 8, and 9 of Appendix A.1.1 in \cite{appendix}. When there is no noise, all three sampling methods offer similar accuracy, and no differences in solution plots are observed in Fig. 7 of Appendix A.1.1 in \cite{appendix}. For a noisy measurement case, the uniform sampling method gives the lowest relative error. 

\begin{figure}[h]
    \centering
    \subfigure[Average Relative Error (\%)]{
        \includegraphics[width=0.45\textwidth]{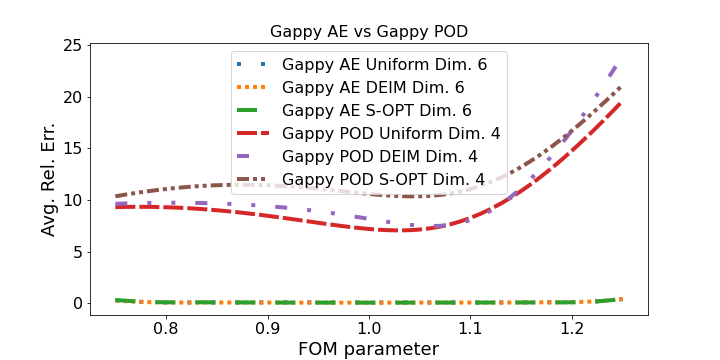}}
    \subfigure[Maximum Relative Error (\%)]{
        \includegraphics[width=0.45\textwidth]{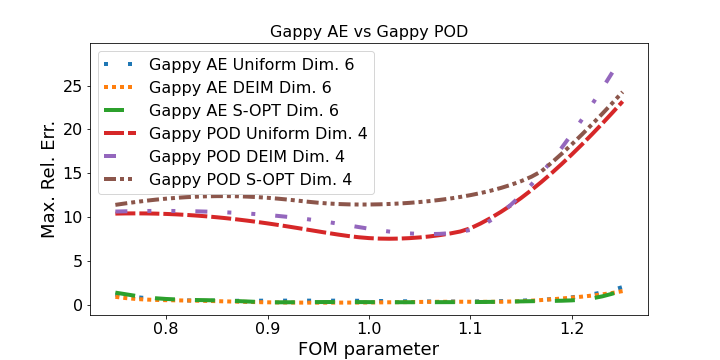}}
    \caption{Gappy AE vs. Gappy POD for diffusion problem with noiseless measurements on the boundary}
\label{fg:ex16BndryGappyAEvsGappyPOD}
\end{figure}

\begin{figure}[h]
    \centering
    \includegraphics[width=0.9\textwidth]{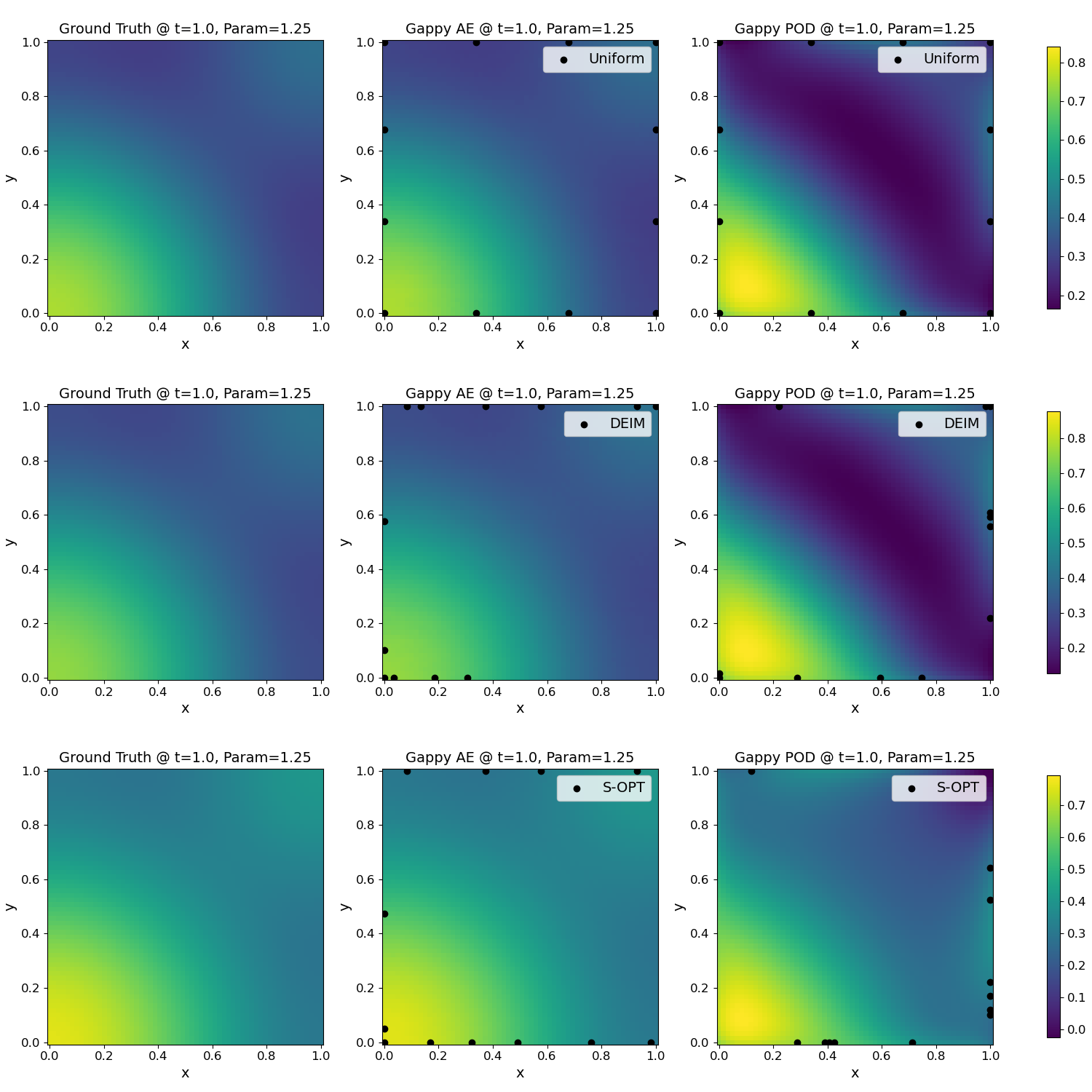}
    \caption{Gappy AE vs. Gappy POD solutions for diffusion problem with noiseless measurements on the boundary. Black dots denote sampling points}
\label{fg:ex16BndryGappyAEvsGappyPODSol}
\end{figure}

\begin{figure}[h]
    \centering
    \subfigure[Average Relative Error (\%)]{
        \includegraphics[width=0.45\textwidth]{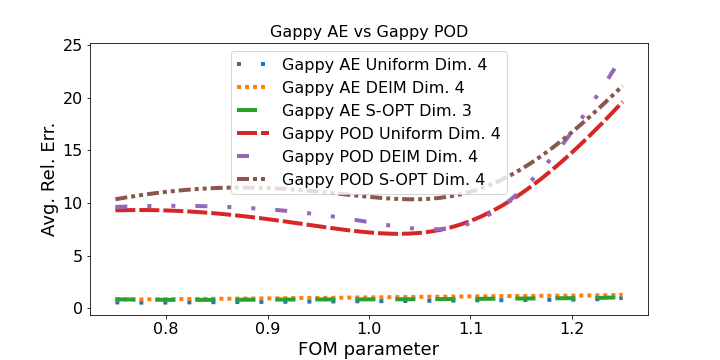}}
    \subfigure[Maximum Relative Error (\%)]{
        \includegraphics[width=0.45\textwidth]{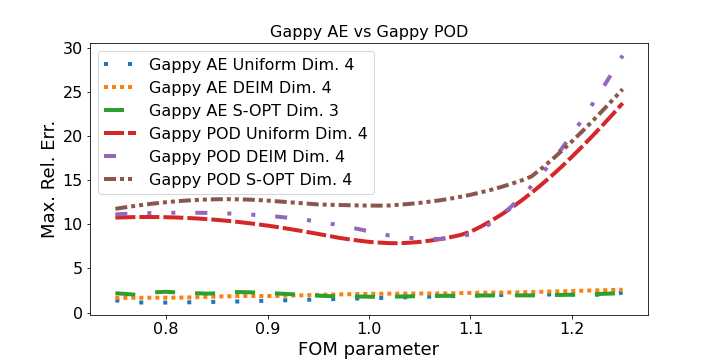}}
    \caption{Gappy AE vs. Gappy POD for diffusion problem with noisy measurements on the boundary}
\label{fg:ex16BndryNoisyGappyAEvsGappyPOD}
\end{figure}

\begin{figure}[h]
    \centering
    \includegraphics[width=0.9\textwidth]{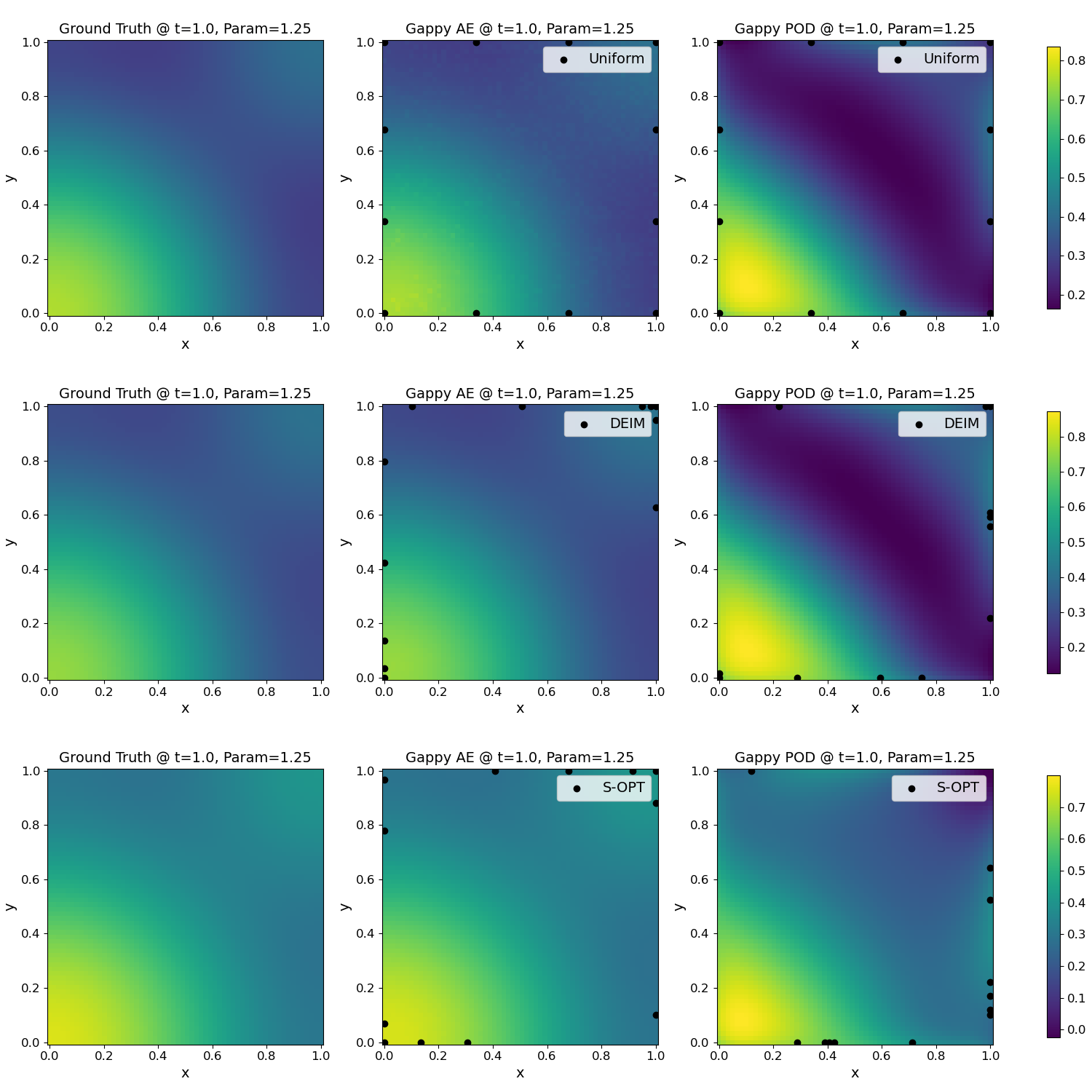}
    \caption{Gappy AE vs. Gappy POD solutions for diffusion problem with noisy measurements on the boundary. Black dots denote sampling points}
\label{fg:ex16BndryNoisyGappyAEvsGappyPODSol}
\end{figure}

Finally, Gappy AE and POD methods are compared in Figs. \ref{fg:ex16BndryGappyAEvsGappyPOD}, \ref{fg:ex16BndryGappyAEvsGappyPODSol}, \ref{fg:ex16BndryNoisyGappyAEvsGappyPOD}, and \ref{fg:ex16BndryNoisyGappyAEvsGappyPODSol}. Regardless of noise in the measurement data, the Gappy AE algorithm outperforms the Gappy POD in terms of accuracy.
 \clearpage

 \clearpage

 \clearpage

 \clearpage

\subsubsection{Sensor Placement in Domain}\label{sec:diffusion_inner}
This section explores scenarios where sensors are placed within the inner domain region. The study varies the auto-encoder's latent space dimension between 3 and 6, employing four different sampling methods: uniform sampling, DEIM, S-OPT, and LHS. 

\begin{figure}[h]
  \centering
  \subfigure[Average Relative Error (\%)]{
  \includegraphics[width=0.9\textwidth]{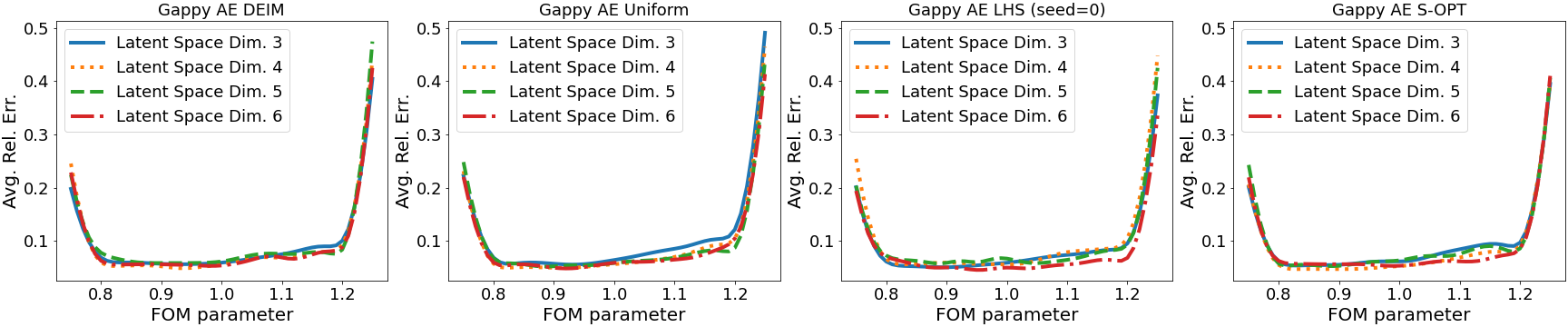}}
  \subfigure[Maximum Relative Error (\%)]{
  \includegraphics[width=0.9\textwidth]{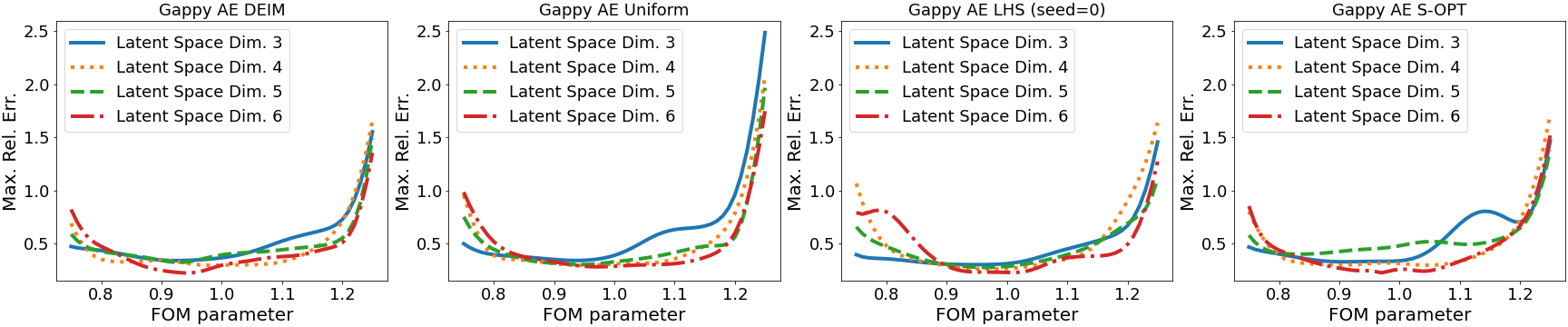}}
  \caption{Accuracy of Gappy AE for diffusion problem with noiseless measurements in the domain} 
  \label{fg:ex16InnerGappyAESize}
\end{figure}

\begin{figure}[h]
  \centering
  \subfigure[Average Relative Error (\%)]{
  \includegraphics[width=0.9\textwidth]{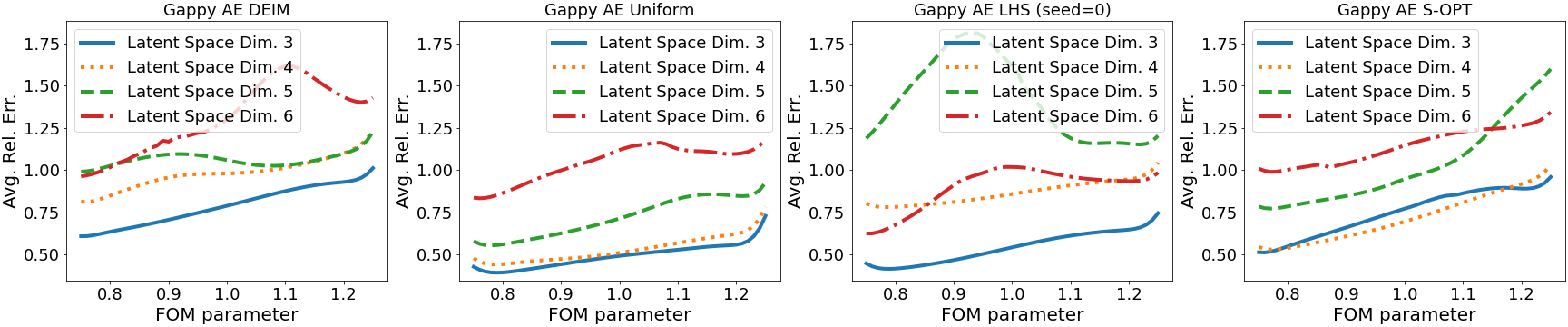}}
  \subfigure[Maximum Relative Error (\%)]{
  \includegraphics[width=0.9\textwidth]{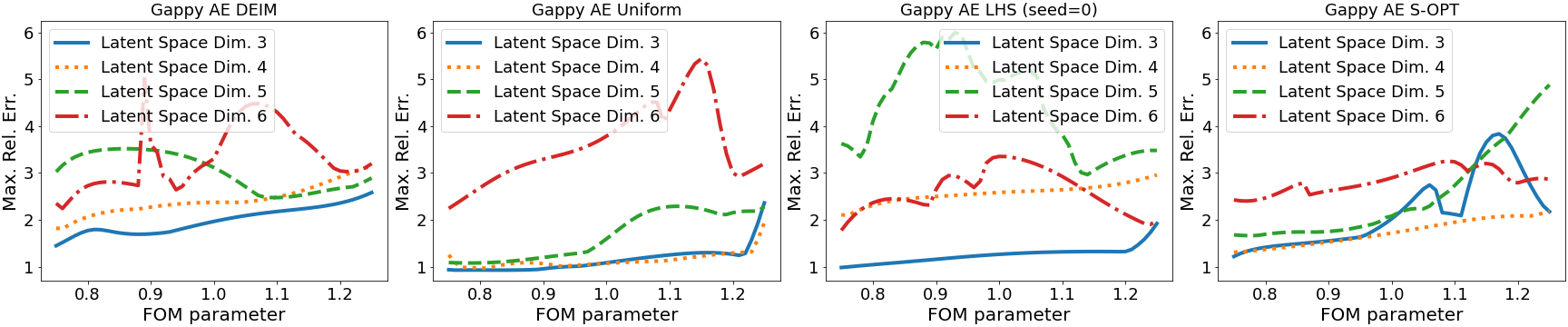}}
  \caption{Accuracy of Gappy AE for diffusion problem with noisy measurements in the domain} 
  \label{fg:ex16InnerNoisyGappyAESize}
\end{figure}

The effects of the latent space dimension in Fig. \ref{fg:ex16InnerGappyAESize} demonstrate no significant changes in relative errors with increasing dimension when there is no noise in measurement data. However, with measurement noise, as shown in Fig. \ref{fg:ex16InnerNoisyGappyAESize}, Gappy AE exhibits better accuracy with a smaller latent space dimension, which is due to the less sensitivity of its auto-encoder to the noise.
 \clearpage

 \clearpage

The section also includes a comparison of the Gappy AE performance for each sampling method. The results in Figs. 10, 11, 12, and 13 of Appendix A.1.2 in \cite{appendix} indicate similar results across all methods with no noise. For a noisy measurement case described in Section \ref{sec:results},
the uniform sampling method yields the most accurate results. Adding noise to measurement data not only deteriorates the data reconstruction accuracy but also weakens the spatial characteristics of the dynamics embedded in the data. Thus, the uniform sampling method works the best among the 4 different sampling methods.

\begin{figure}[h]
    \centering
    \subfigure[Average Relative Error (\%)]{
        \includegraphics[width=0.45\textwidth]{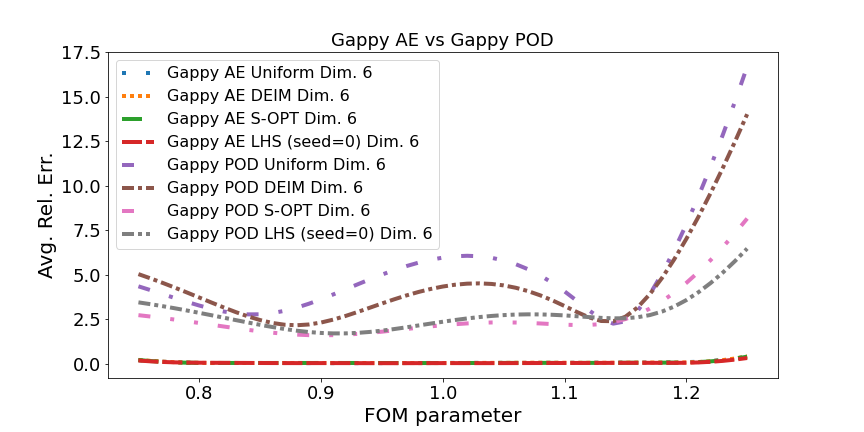}}
    \subfigure[Maximum Relative Error (\%)]{
        \includegraphics[width=0.45\textwidth]{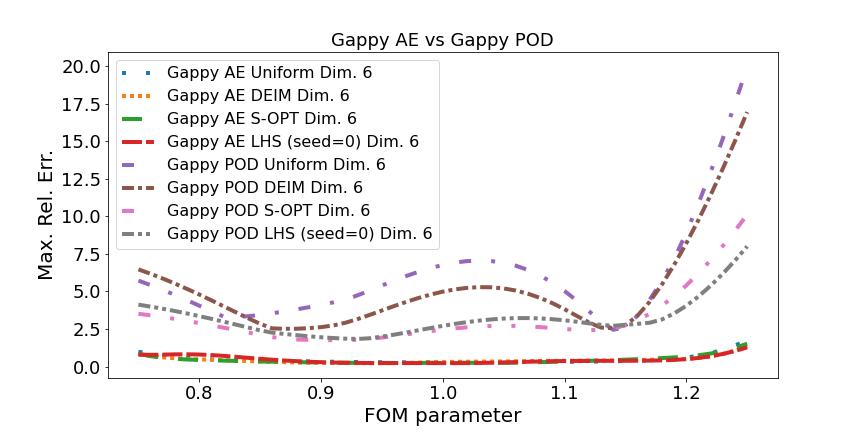}}
    \caption{Gappy AE vs. Gappy POD for diffusion problem with noiseless measurements in the domain} 
\label{fg:ex16InnerGappyAEvsGappyPOD}
\end{figure}

\begin{figure}[h]
    \centering
    \includegraphics[width=0.7\textwidth]{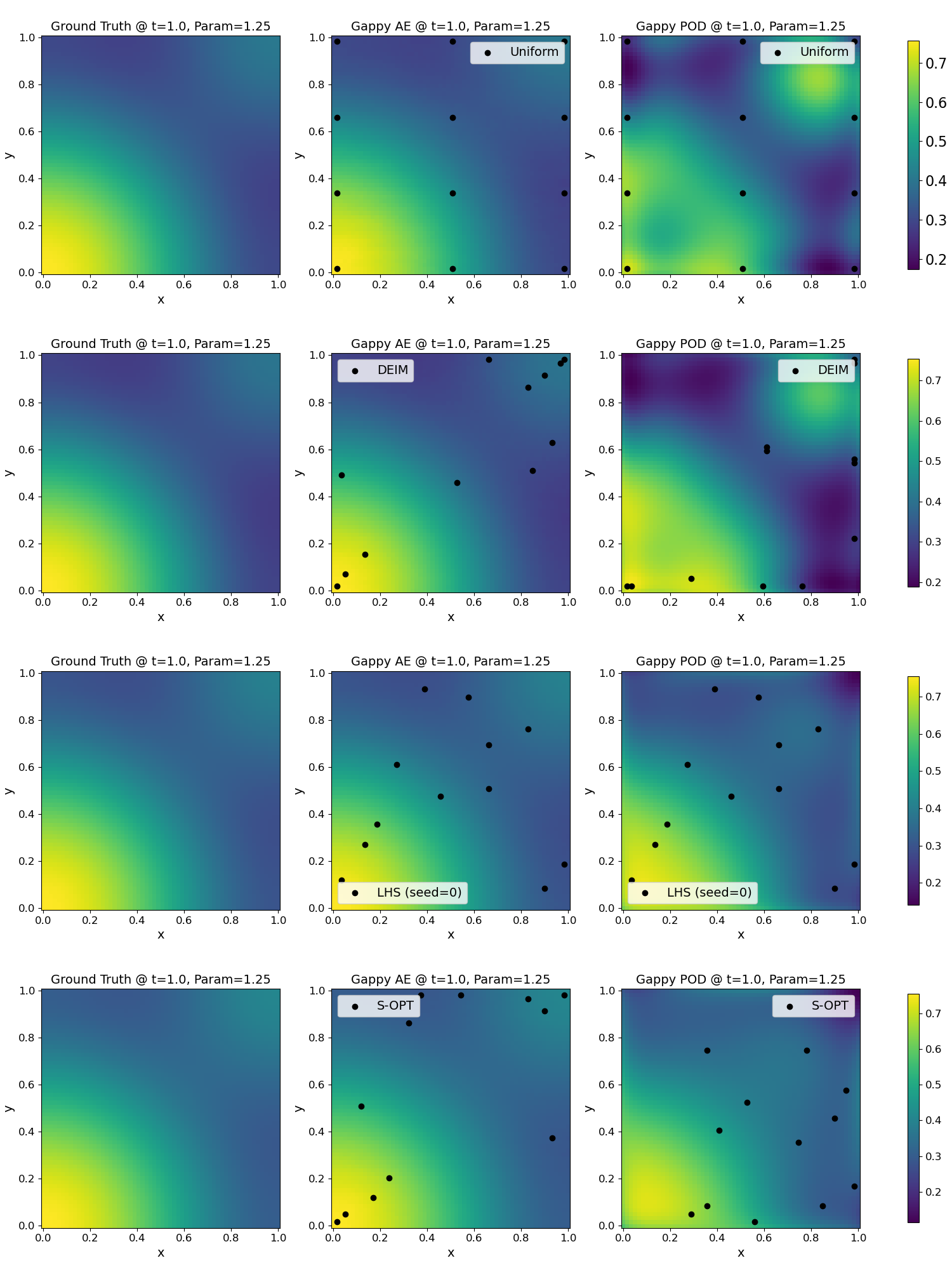}
    \caption{Gappy AE vs. Gappy POD solutions for diffusion problem with noiseless measurements in the domain. Black dots denote sampling points}
\label{fg:ex16InnerGappyAEvsGappyPODSol}
\end{figure}

\begin{figure}[h]
    \centering
    \subfigure[Average Relative Error (\%)]{
        \includegraphics[width=0.45\textwidth]{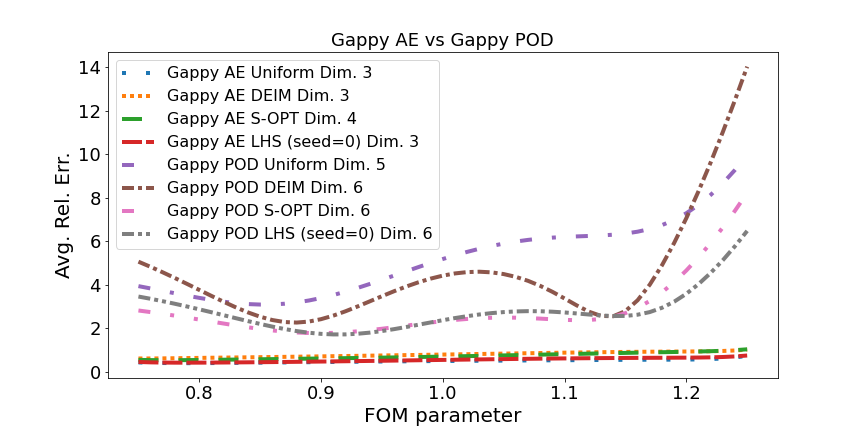}}
    \subfigure[Maximum Relative Error (\%)]{
        \includegraphics[width=0.45\textwidth]{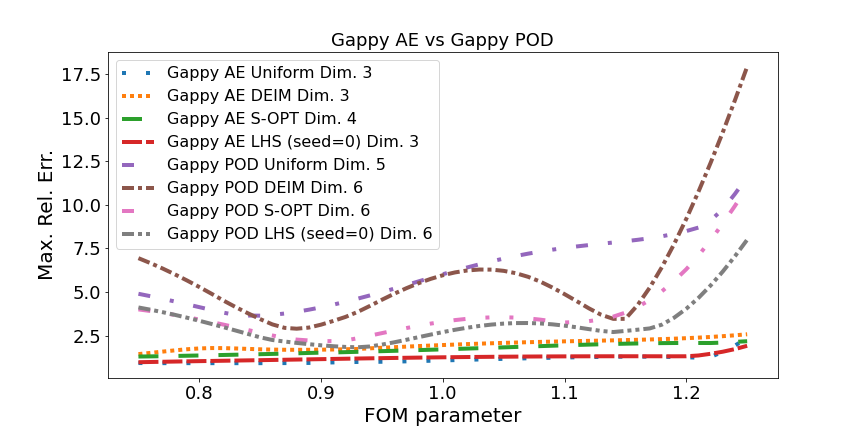}}
    \caption{Gappy AE vs. Gappy POD for diffusion problem with noisy measurements in the domain}
\label{fg:ex16InnerNoisyGappyAEvsGappyPOD}
\end{figure}

\begin{figure}[h]
    \centering
    \includegraphics[width=0.7\textwidth]{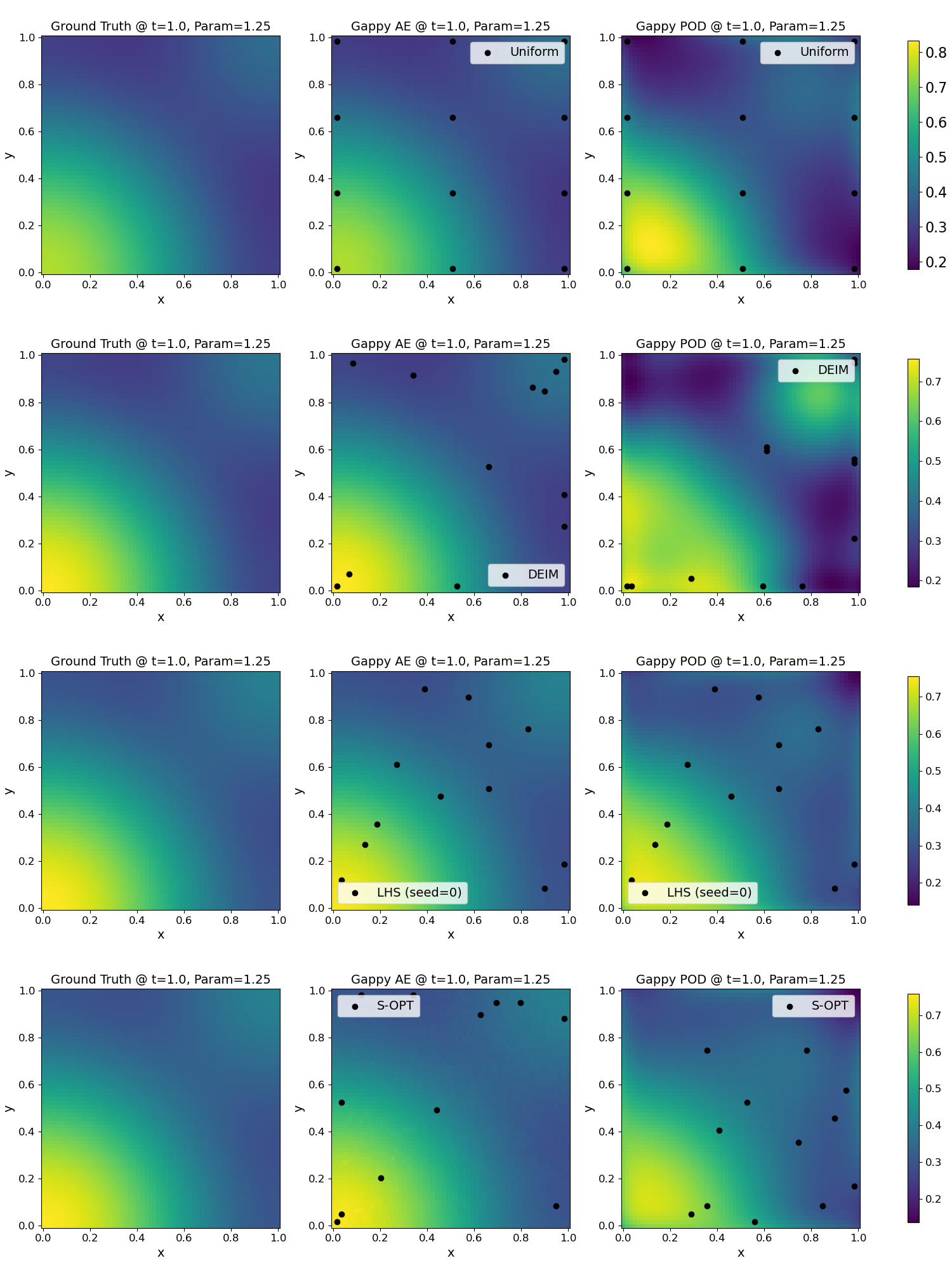}
    \caption{Gappy AE vs. Gappy POD solutions for diffusion problem with noisy measurements in the domain. Black dots denote sampling points}
\label{fg:ex16InnerNoisyGappyAEvsGappyPODSol}
\end{figure}

Finally, Gappy AE and Gappy POD methods are compared in Figs. \ref{fg:ex16InnerGappyAEvsGappyPOD}, \ref{fg:ex16InnerGappyAEvsGappyPODSol}, \ref{fg:ex16InnerNoisyGappyAEvsGappyPOD}, and \ref{fg:ex16InnerNoisyGappyAEvsGappyPODSol}. Regardless of noise in the measurement data, Gappy AE algorithm outperforms Gappy POD in terms of accuracy.
 \clearpage

 \clearpage

 \clearpage

 \clearpage

\subsection{Radial Advection Problem}\label{sec:advection}
\begin{figure}[h]
  \centering
  \subfigure[$\mu_{min}=0.75$]{
  \includegraphics[width=0.9\textwidth]{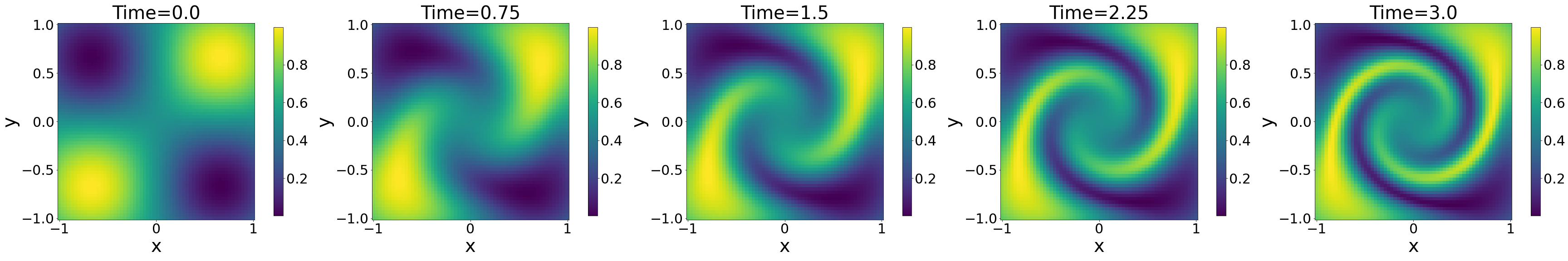}}
  \subfigure[$\mu_{max}=1.25$]{
  \includegraphics[width=0.9\textwidth]{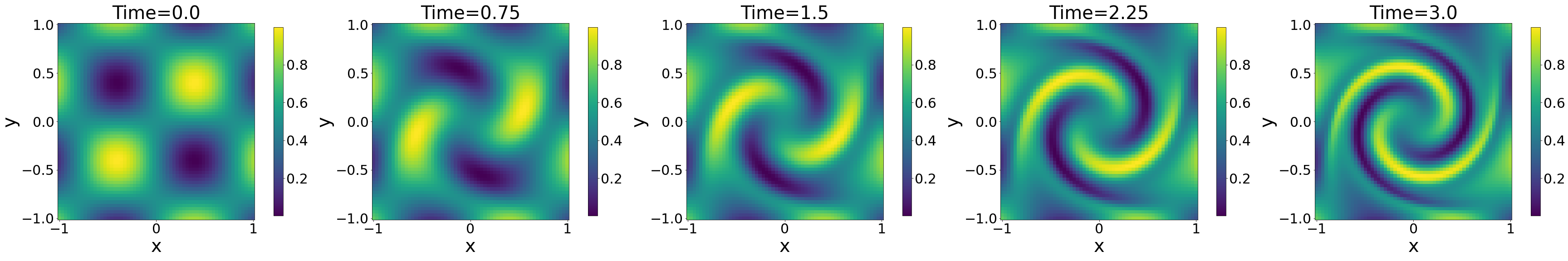}}
  \caption{Radial advection simulation solutions from the initial to final time for two endpoints of $\mu$}
  \label{fg:ex9SolTwoParam}
\end{figure}

This section considers the following 2D parameterized radial advection problem: 
\begin{equation}
    \frac{\partial u}{\partial t}+\mathbf{v}\cdot\nabla u =0,
\end{equation}
where the spatial and time domains are given by $\mathbf{x} \in \Omega = [-1,1] \times [-1,1]$ and $t \in [0,3]$, respectively; velocity, $\mathbf{v}$ is defined as
\begin{equation}
    \mathbf{v} = \frac{\pi}{2}(1-\mathrm{x}_1^2)^2(1-\mathrm{x}_2^2)^2(\mathrm{x}_2,-\mathrm{x}_1)^T.
\end{equation}
The Dirichlet boundary condition
\begin{equation}
    u=0 \quad \text{on} \quad \partial \Omega
\end{equation}
and initial condition
\begin{equation}
    u(\mathbf{x},0;\mu)=0.5(\sin(\pi\mu \mathrm{x}_1)\sin(\pi\mu \mathrm{x}_2)+1)
\end{equation}
are imposed, where $\mu$ is a parameter. The spatial domain is discretized into a $64\times64$ square mesh. The fourth-order Runge-Kutta time stepping scheme is applied with uniform $\Delta t=0.005$. This problem is based on the example 9 of MFEM's example problems. The source code of the radial advection problem is available at \url{https://github.com/mfem/mfem/blob/master/examples/ex9.cpp}. Figure~\ref{fg:ex9SolTwoParam} shows five snapshots for the two extreme parameter values.

For this radial advection problem, we impose the Dirichlet boundary condition, and its value is constant. Thus, we only allow sensors to be located in the inner domain region because data measured at the boundary never changes.

 \clearpage

\subsubsection{Sensor Placement in Domain}\label{sec:advection_inner}
This section consider placing sensors exclusively within the inner domain region. The auto-encoder's latent space dimension ranges from 3 to 6, and four sampling methods are utilized, i.e., uniform sampling, DEIM, S-OPT, and LHS.

\begin{figure}[h]
  \centering
  \subfigure[Average Relative Error (\%)]{
  \includegraphics[width=0.9\textwidth]{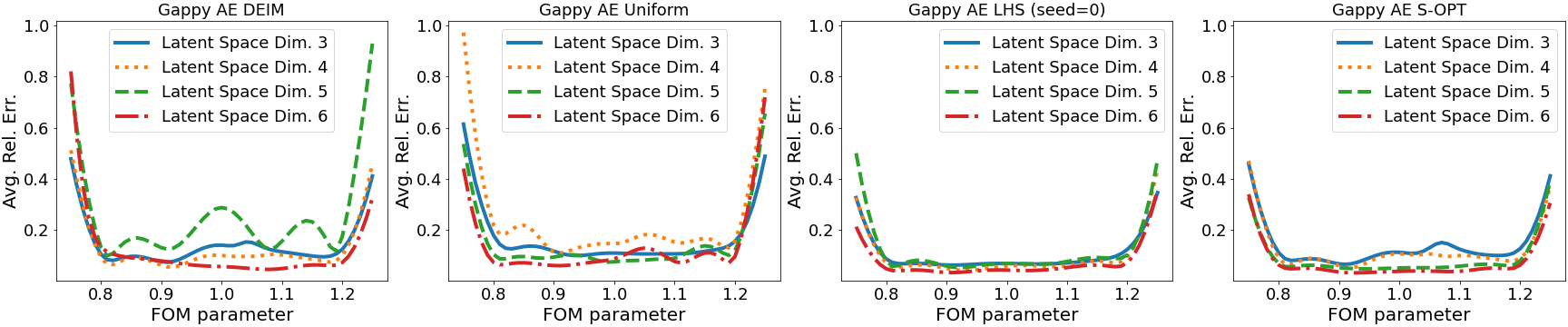}}
  \subfigure[Maximum Relative Error (\%)]{
  \includegraphics[width=0.9\textwidth]{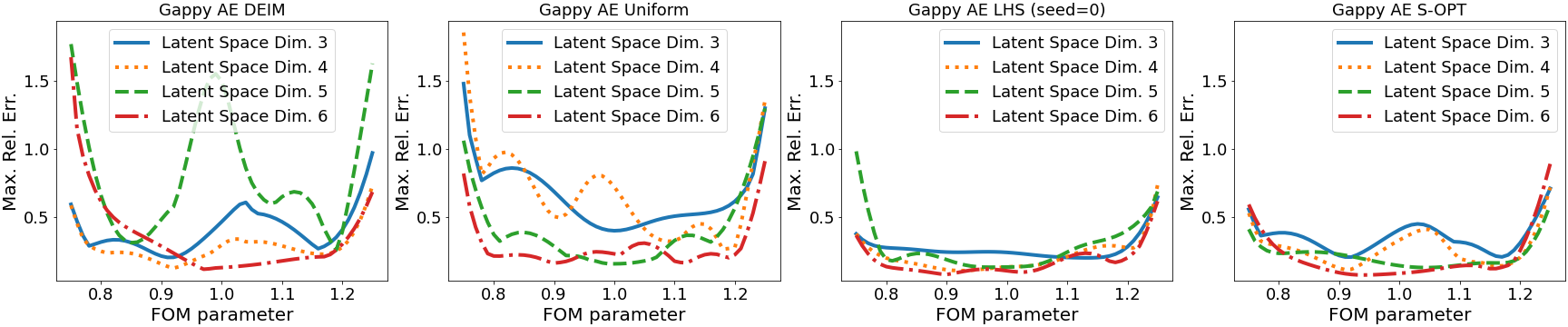}}
  \caption{Accuracy of Gappy AE for radial advection problem with noiseless measurements in the domain}
  \label{fg:ex9InnerGappyAESize}
\end{figure}

\begin{figure}[h]
  \centering
  \subfigure[Average Relative Error (\%)]{
  \includegraphics[width=0.9\textwidth]{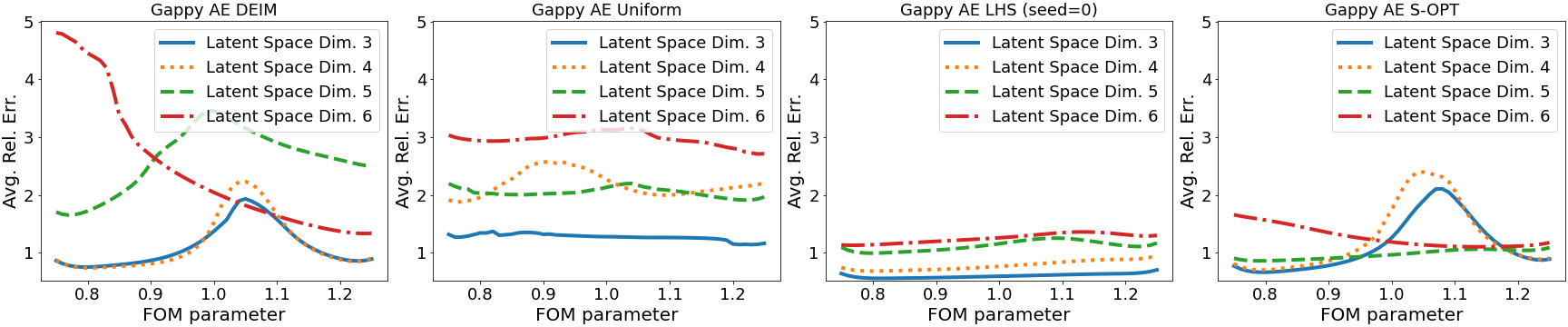}}
  \subfigure[Maximum Relative Error (\%)]{
  \includegraphics[width=0.9\textwidth]{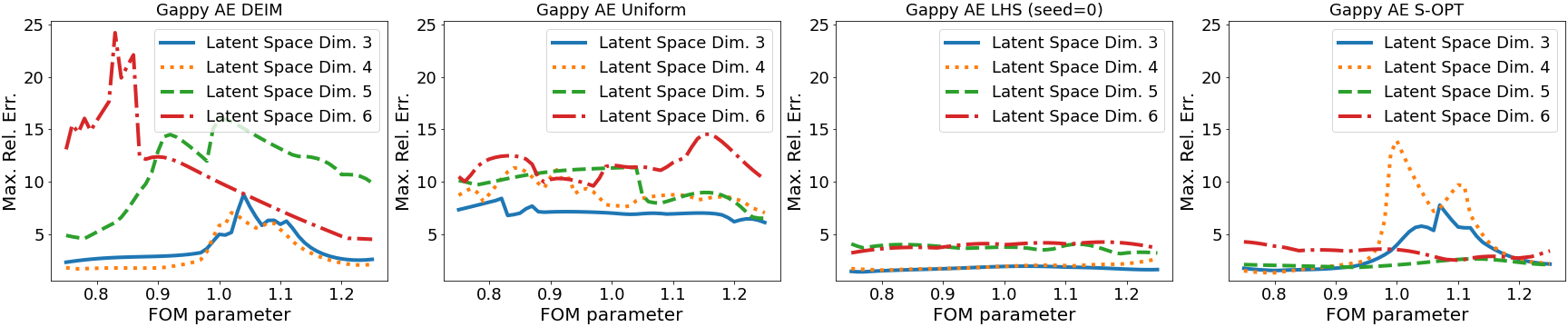}}
  \caption{Accuracy of Gappy AE for radial advection problem with noisy measurements in the domain}
  \label{fg:ex9InnerNoisyGappyAESize}
\end{figure}

The findings indicate that relative errors are the lowest for all sampling algorithms when the dimension of the latent space is 6, as in Fig. \ref{fg:ex9InnerGappyAESize}. However, with noise, the Gappy AE demonstrates higher accuracy with a smaller latent space dimension (Fig. \ref{fg:ex9InnerNoisyGappyAESize}). Thus, DEIM, uniform sampling, LHS, and S-OPT algorithms work best when the latent space dimension is 4, 3, 3, and 5, respectively.

 \clearpage

 \clearpage

Next, we choose the best-performing latent space dimension for each sampling method and compared them in Figs. 14, 15, 16, and 17 of Appendix A.2.1 in \cite{appendix}. The results reveal similar results for S-OPT and LHS methods with no noise. However, the accuracy of the DEIM sampling method deteriorates at parameters near 0.8. The uniform sampling method yields less accurate results overall. In contrast, the overall accuracy decreases with noise. The LHS method gives the lowest relative error with noisy data among different sampling algorithms.

\begin{figure}[h]
    \centering
    \subfigure[Average Relative Error (\%)]
    {\includegraphics[width=0.45\textwidth]{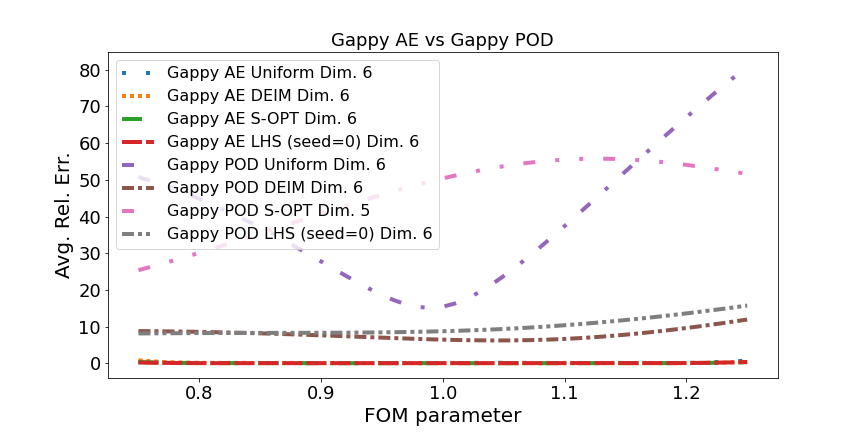}}
    \subfigure[Maximum Relative Error (\%)]
      {\includegraphics[width=0.45\textwidth]{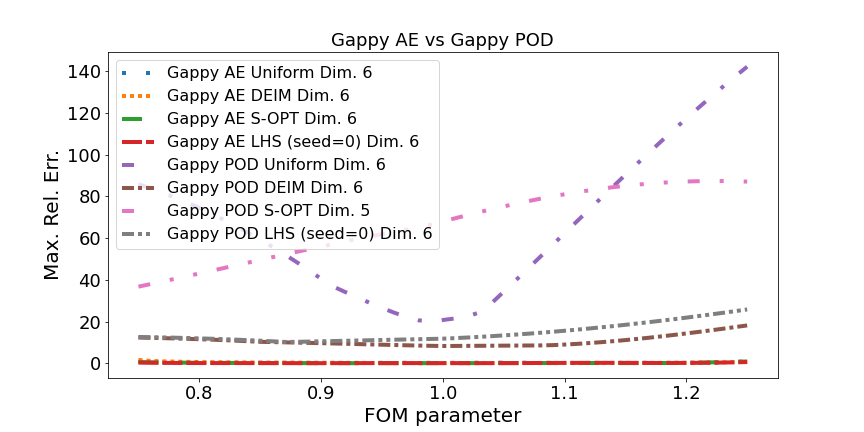}}
     \caption{Gappy AE vs. Gappy POD for radial advection problem with noiseless measurements in the domain}
\label{fg:ex9InnerGappyAEvsGappyPOD}
\end{figure}

\begin{figure}[h]
    \centering
    \includegraphics[width=0.7\textwidth]{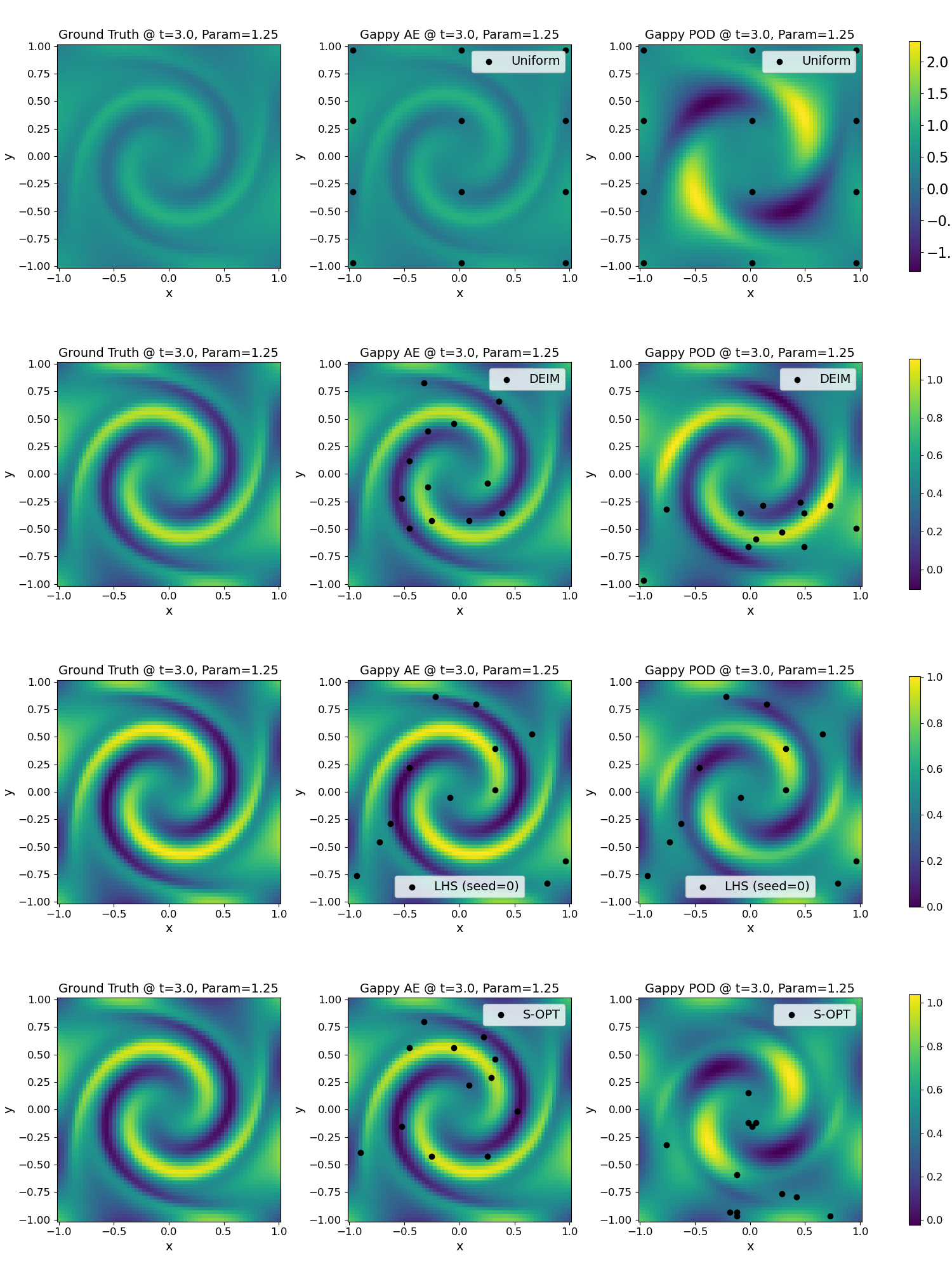}
    \caption{Gappy AE vs. Gappy POD solutions for radial advection problem with noiseless measurements. Black dots denote sampling points}
\label{fg:ex9InnerGappyAEvsGappyPODSol}
\end{figure}

\begin{figure}[h]
    \centering
    \subfigure[Average Relative Error (\%)]
     {\includegraphics[width=0.45\textwidth]{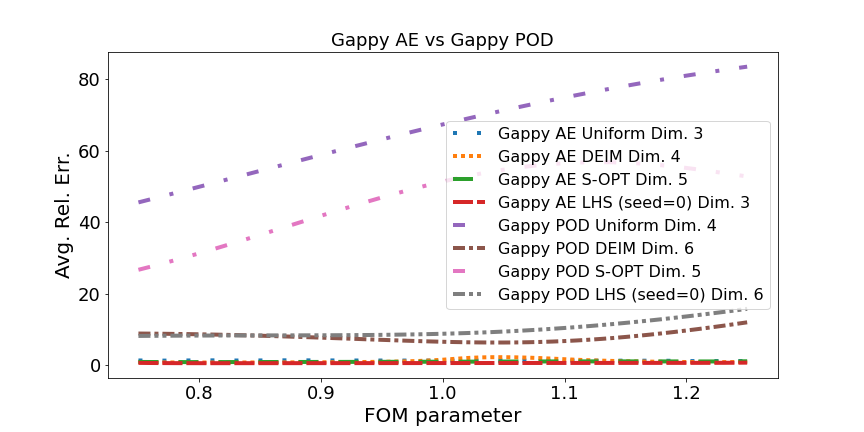}}
    \subfigure[Maximum Relative Error (\%)]
      {\includegraphics[width=0.45\textwidth]{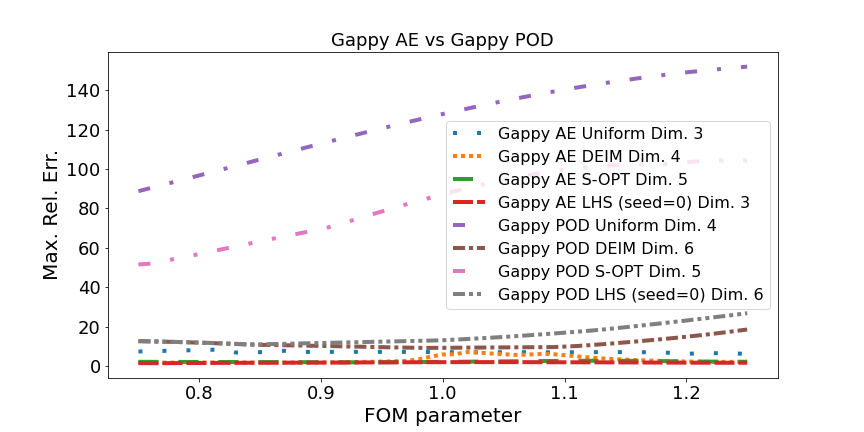}}
      \caption{Gappy AE vs. Gappy POD for radial advection problem with noisy measurements in the domain}
\label{fg:ex9InnerNoisyGappyAEvsGappyPOD}
\end{figure}

\begin{figure}[h]
    \centering
    \includegraphics[width=0.7\textwidth]{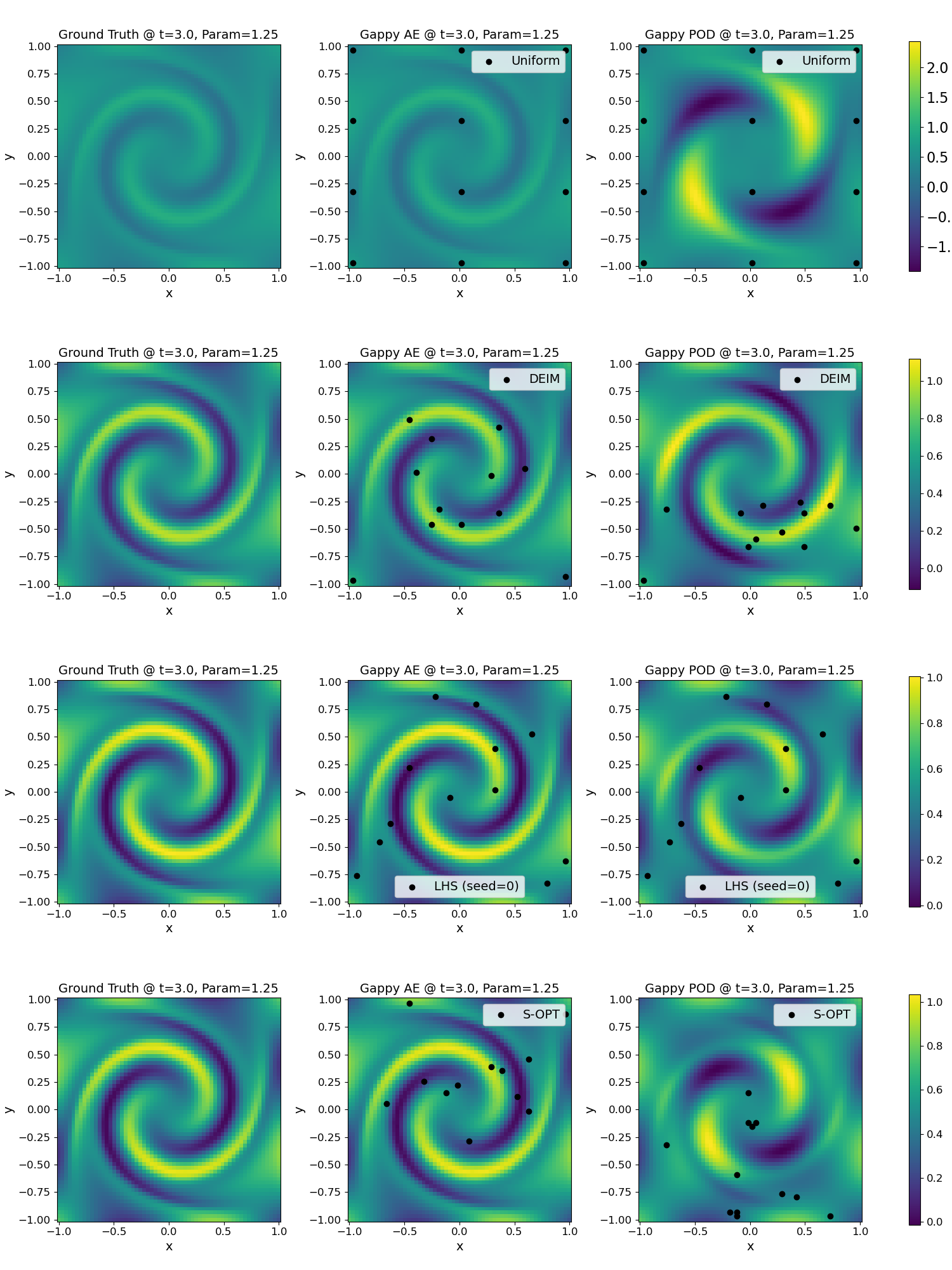}
    \caption{Gappy AE vs. Gappy POD solutions for radial advection problem with noisy measurements in the domain. Black dots denote sampling points}
\label{fg:ex9InnerNoisyGappyAEvsGappyPODSol}
\end{figure}

A comparison is made between Gappy AE and Gappy POD methods, as illustrated in Figs. \ref{fg:ex9InnerGappyAEvsGappyPOD}, \ref{fg:ex9InnerGappyAEvsGappyPODSol}, \ref{fg:ex9InnerNoisyGappyAEvsGappyPOD}, and \ref{fg:ex9InnerNoisyGappyAEvsGappyPODSol}. The Gappy AE consistently demonstrates superior accuracy, irrespective of noise in the measurement data. The results for the Gappy POD can be found in Appendix B of \cite{appendix}.
 \clearpage

 \clearpage

 \clearpage

 \clearpage

\subsection{Wave Problem}\label{sec:wave}
\begin{figure}[h]
  \centering
  \subfigure[$\mu_{min}=0.75$]{
  \includegraphics[width=0.9\textwidth]{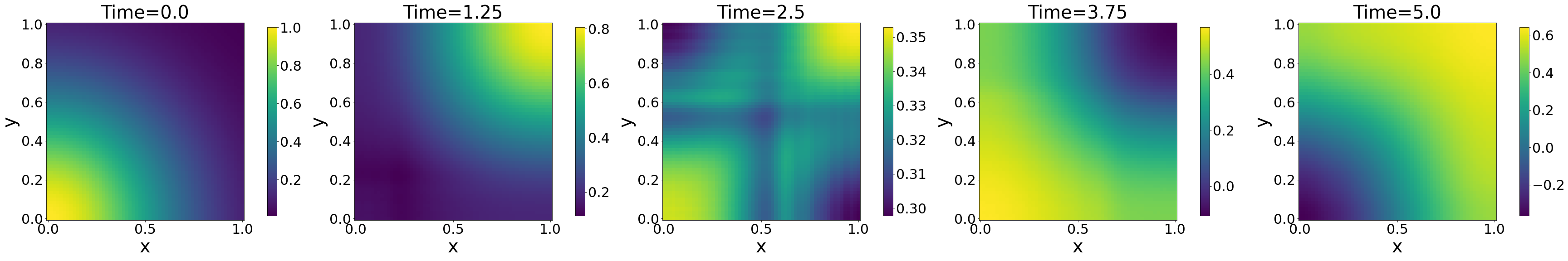}}
  \subfigure[$\mu_{max}=1.25$]{
  \includegraphics[width=0.9\textwidth]{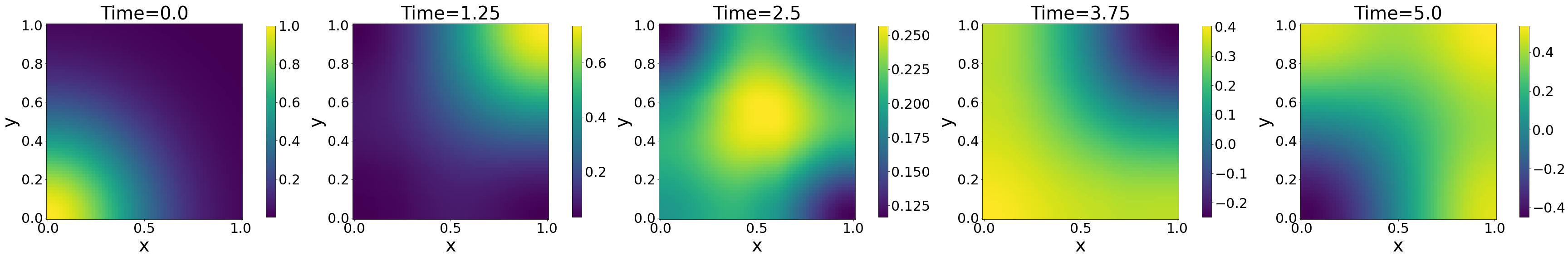}}
  \caption{Wave simulation solutions from the initial to the final time for two endpoints of $\mu$}
  \label{fg:ex23SolTwoParam}
\end{figure}

A 2D parameterized wave equation is considered
\begin{equation}
    \frac{\partial^2 u}{\partial t^2} - \mu ^2 \nabla u = 0,
\end{equation}
where $\mu$ is a parameter. The spatial and time domains are set as $\mathbf{x}\in[0,1]\times[0,1]$ and $t \in [0,5]$, respectively. The Neumann boundary condition is imposed
\begin{equation}
    \frac{\partial u}{\partial \mathbf{x}}\cdot\mathbf{n}=0 \quad \text{on} \quad \partial \Omega,
\end{equation}
where $\mathbf{n}$ denotes the unit normal vector. The initial condition
\begin{equation}
    u(\mathbf{x},0) = \exp{(-3\|\mathbf{x}\|_2^2)}
\end{equation}
is imposed. Discretizing the spatial domain into a $64\times64$ square mesh gives us a uniform mesh grid. The time domain is discretized with uniform $\Delta t=0.01$. The generalized alpha scheme with $\rho=1$ is employed for the time integrator. This wave problem is example 23 among MFEM's demo examples. The source code of the wave problem is available at \url{https://github.com/mfem/mfem/blob/master/examples/ex23.cpp}. Figure~\ref{fg:ex23SolTwoParam} shows five snapshots for two extreme parameter values.

As in the diffusion problem case, two sensor location scenarios are considered: boundaries and inner regions of the domain.

 \clearpage

\subsubsection{Sensor Placement on Boundary}\label{sec:wave_bndry}
This section considers scenarios where sensors are positioned at the boundary region of the domain. The study explores varying the latent space dimension of the auto-encoder between 3 and 6, employing three different sampling methods: uniform sampling, DEIM, and S-OPT. 

\begin{figure}[h]
  \centering
  \subfigure[Average Relative Error (\%)]{
  \includegraphics[width=0.9\textwidth]{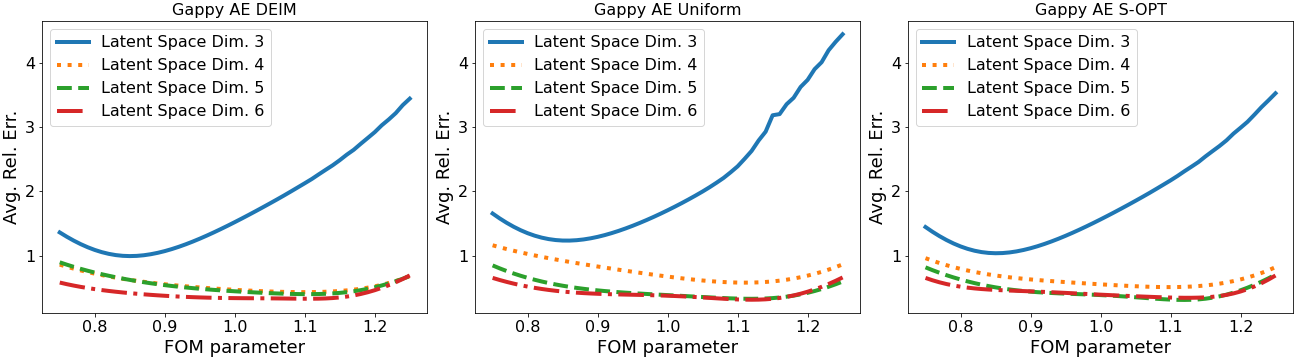}}
  \subfigure[Maximum Relative Error (\%)]{
  \includegraphics[width=0.9\textwidth]{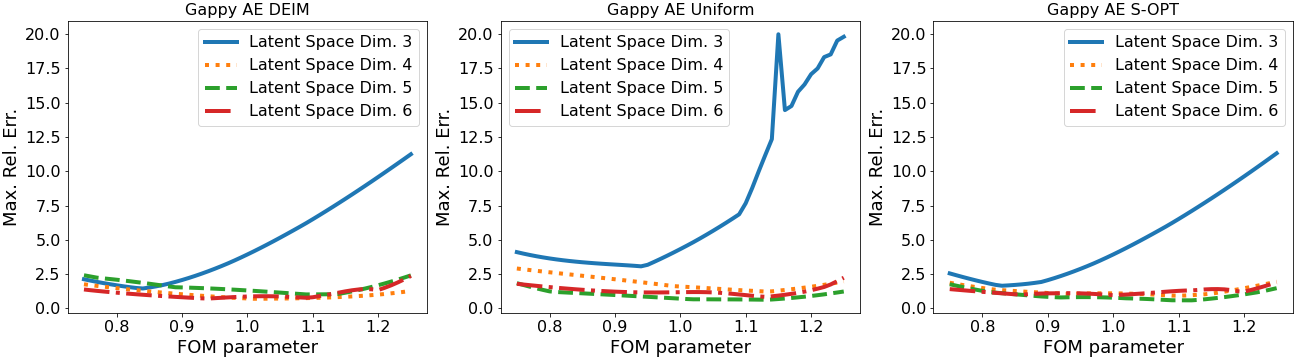}}
  \caption{Accuracy of Gappy AE for wave problem with noiseless measurements on the boundary} 
  \label{fg:ex23BndryGappyAESize}
\end{figure}

\begin{figure}[h]
  \centering
  \subfigure[Average Relative Error (\%)]{
  \includegraphics[width=0.9\textwidth]{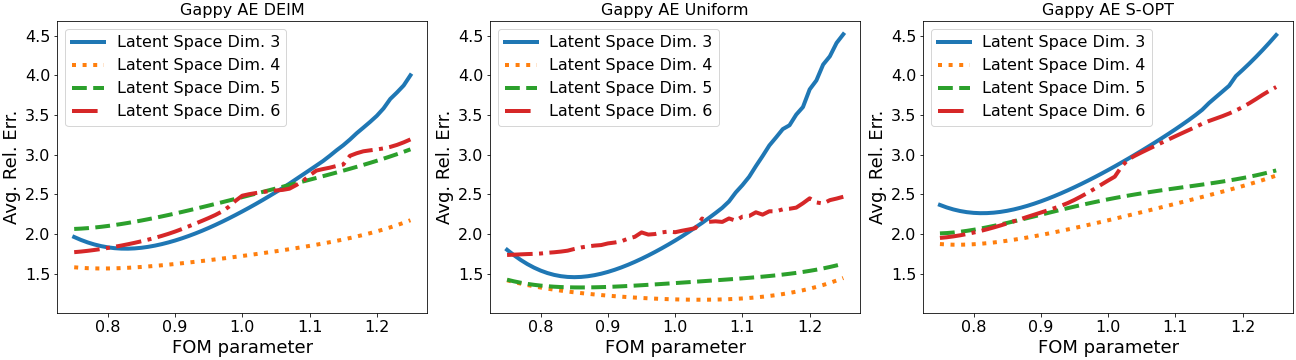}}
  \subfigure[Maximum Relative Error (\%)]{
  \includegraphics[width=0.9\textwidth]{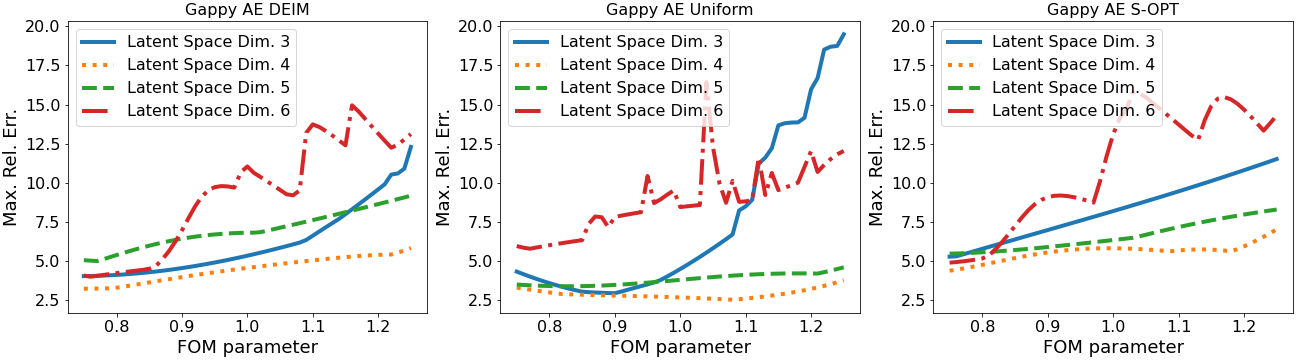}}
  \caption{Accuracy of Gappy AE for wave problem with noisy measurements on the boundary}
  \label{fg:ex23BndryNoisyGappyAESize}
\end{figure}

The results in Figs. \ref{fg:ex23BndryGappyAESize} and \ref{fg:ex23BndryNoisyGappyAESize} reveal that for 3 different sampling algorithms, data reconstruction accuracies of the Gappy AE with noiseless measurement are similar for latent space dimensions of 4, 5, and 6. When the latent space dimension is 3, the relative errors are worst for all sampling cases. However, noise in the measurement data leads to different results. The Gappy AE with the latent space dimension of 4 gives us the lowest relative error among other latent space dimensions for three different sampling algorithms.
 \clearpage

 \clearpage

The comparison of the data reconstruction error of Gappy AE with the latent space that gives the best performance for each sampling method is shown in Figs. 18, 19, 20, and 21 of Appendix A.3.1 in \cite{appendix}. The Gappy AE shows negligible differences in solution plots in Fig. 19 of Appendix A.3.1 in \cite{appendix} when there is no noise. However, in Fig. 20 of Appendix A.3.1 in \cite{appendix}, the Gappy AE with uniform sampling yields the lowest relative error for noisy measurements. Moreover, adding noise to measurement data aggravates the relative errors.

\begin{figure}[h]
    \centering
    \subfigure[Average Relative Error (\%)]{
        \includegraphics[width=0.45\textwidth]{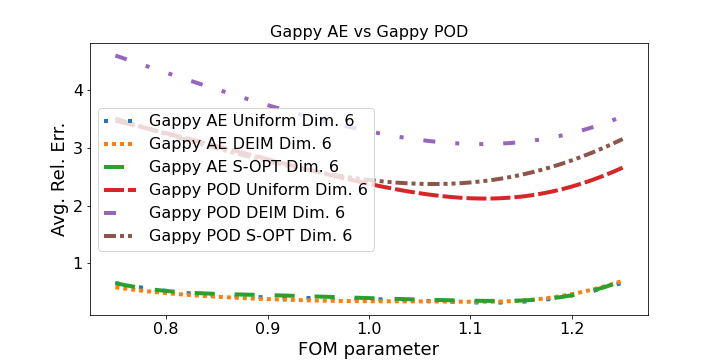}}
    \subfigure[Maximum Relative Error (\%)]{
        \includegraphics[width=0.45\textwidth]{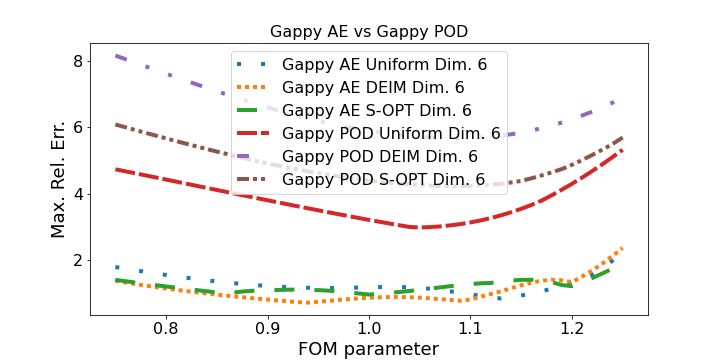}}
    \caption{Gappy AE vs. Gappy POD for wave problem with noiseless measurements on the boundary}
\label{fg:ex23BndryGappyAEvsGappyPOD}
\end{figure}

\begin{figure}[h]
    \centering
    \includegraphics[width=0.9\textwidth]{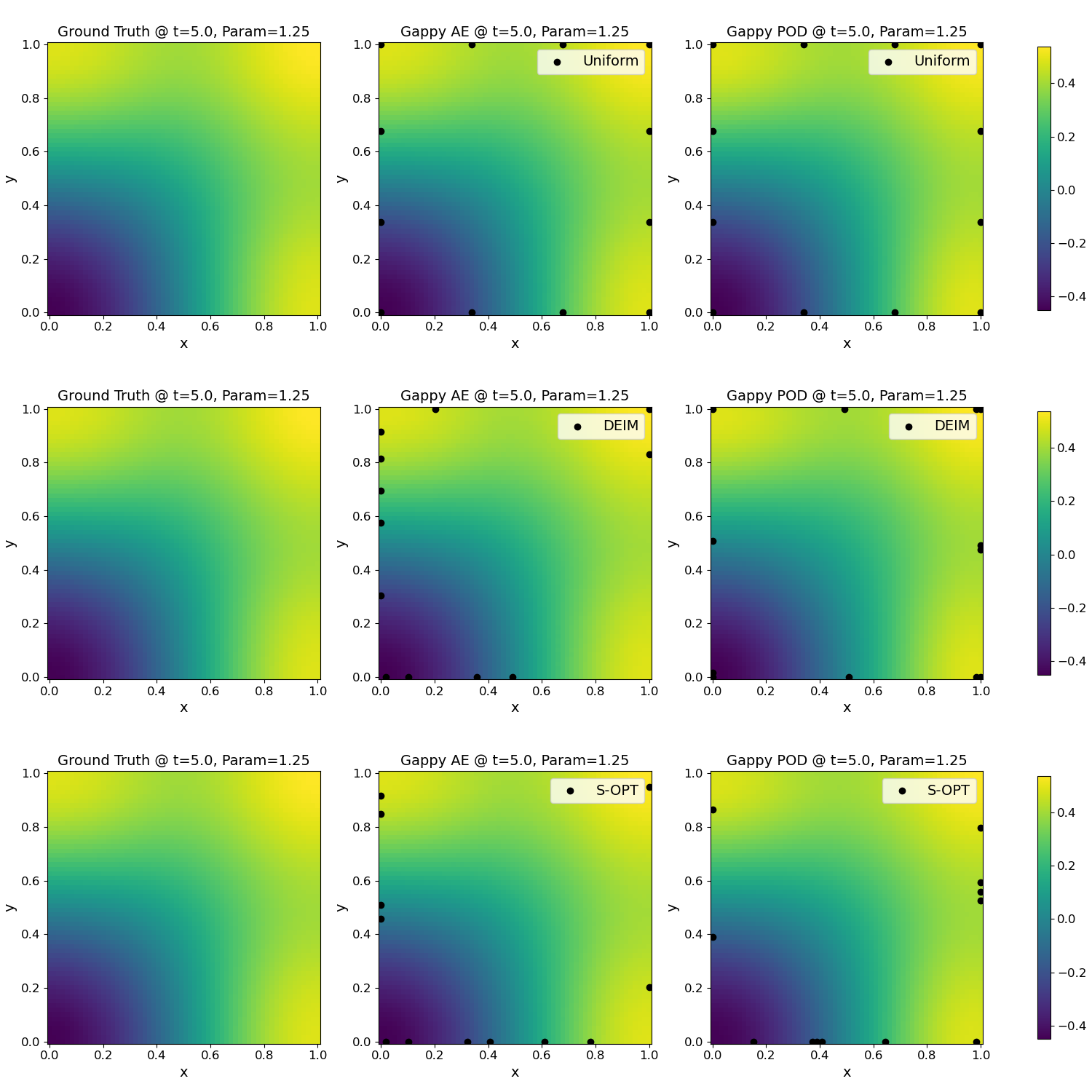}
    \caption{Gappy AE vs. Gappy POD solutions for wave problem with noiseless measurements on the boundary. Black dots denote sampling points}
\label{fg:ex23BndryGappyAEvsGappyPODSol}
\end{figure}

\begin{figure}[h]
    \centering
    \subfigure[Average Relative Error (\%)]{
        \includegraphics[width=0.45\textwidth]{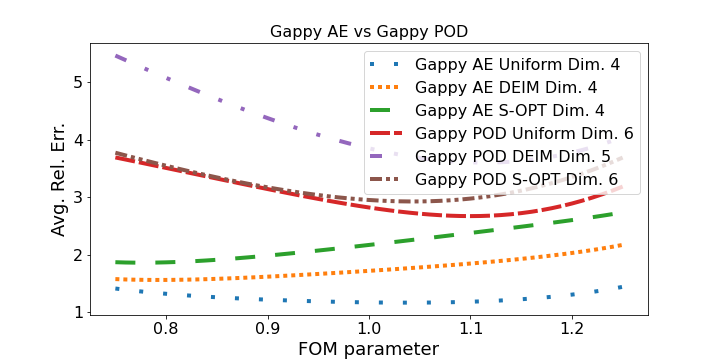}}
    \subfigure[Maximum Relative Error (\%)]{
        \includegraphics[width=0.45\textwidth]{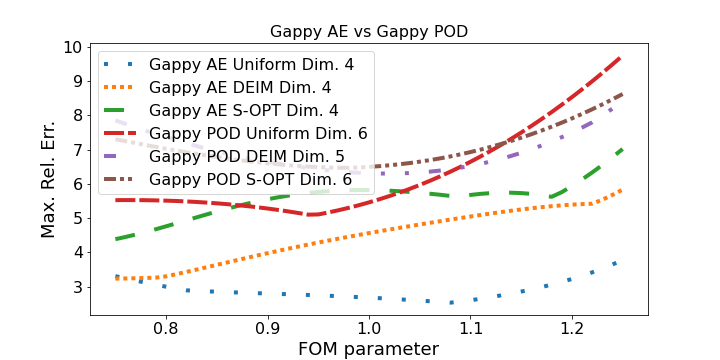}}
    \caption{Gappy AE vs. Gappy POD for wave problem with noisy measurements on the boundary}
\label{fg:ex23BndryNoisyGappyAEvsGappyPOD}
\end{figure}

\begin{figure}[h]
    \centering
    \subfigure[]{
        \includegraphics[width=0.9\textwidth]{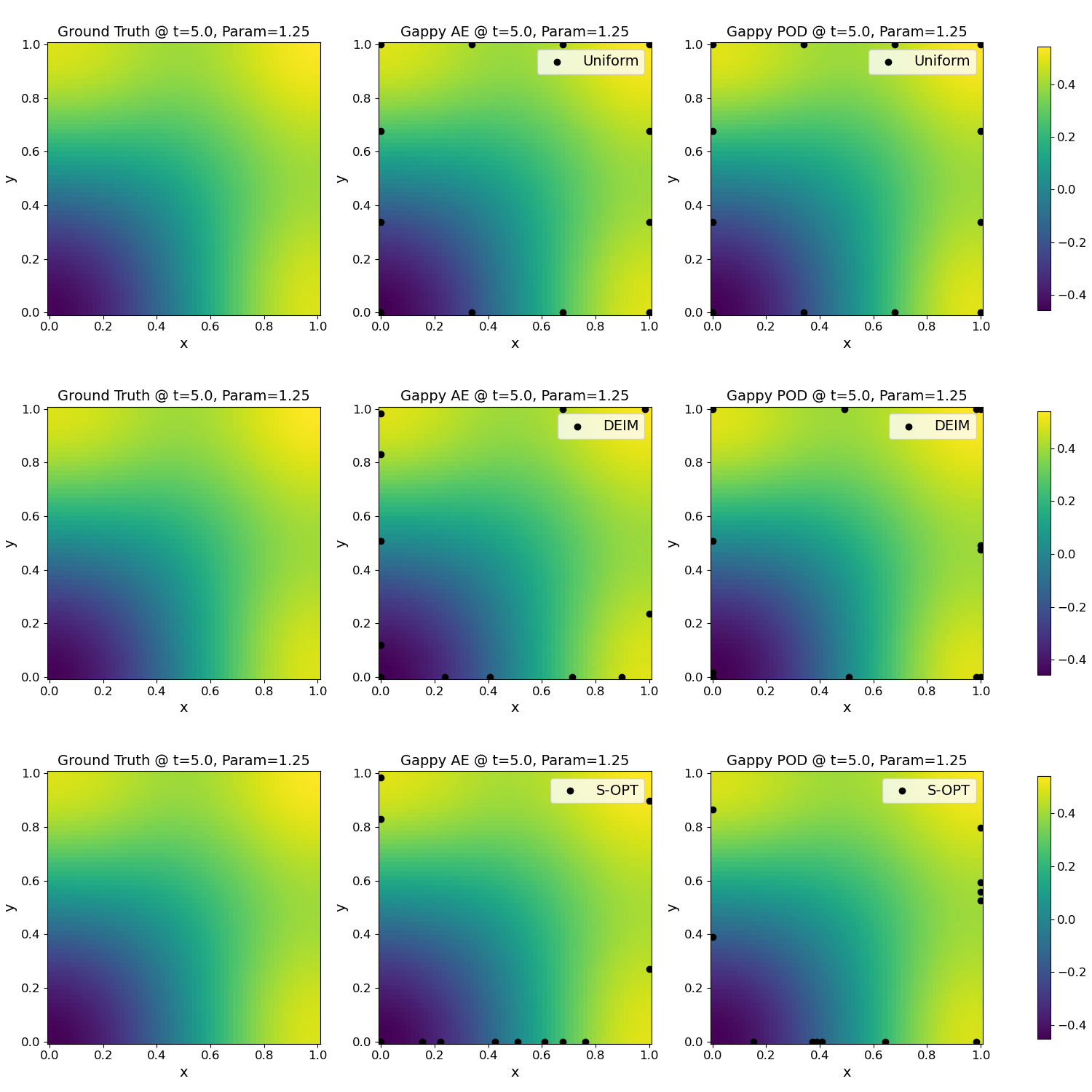}}
    \caption{Gappy AE vs. Gappy POD solutions for wave problem with noisy measurements on the boundary. Black dots denote sampling points}
\label{fg:ex23BndryNoisyGappyAEvsGappyPODSol}
\end{figure}

Finally, Figs. \ref{fg:ex23BndryGappyAEvsGappyPOD}, \ref{fg:ex23BndryGappyAEvsGappyPODSol}, \ref{fg:ex23BndryNoisyGappyAEvsGappyPOD}, and \ref{fg:ex23BndryNoisyGappyAEvsGappyPODSol} compare the Gappy AE and POD methods. Gappy AE consistently outperforms Gappy POD in accuracy, regardless of measurement noise.
 \clearpage

 \clearpage

 \clearpage 

 \clearpage

\subsubsection{Sensor Placement in Domain}\label{sec:wave_inner}
This section assumes that sensors are positioned solely within the inner domain region. The latent space dimension of the auto-encoder is explored within a range of 3 to 6, and four distinct sampling methods are applied: uniform sampling, DEIM, S-OPT, and LHS.

\begin{figure}[h]
  \centering
  \subfigure[Average Relative Error (\%)]{
  \includegraphics[width=0.9\textwidth]{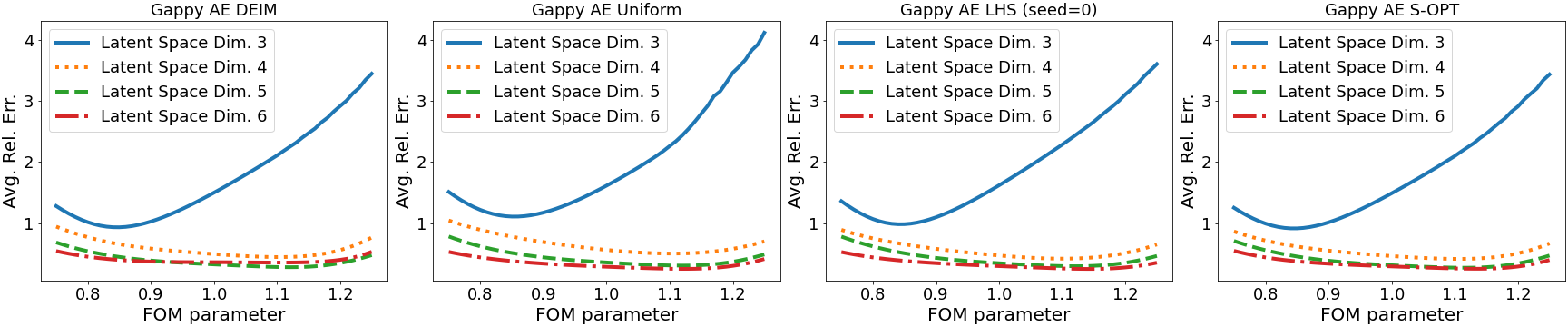}}
  \subfigure[Maximum Relative Error (\%)]{
  \includegraphics[width=0.9\textwidth]{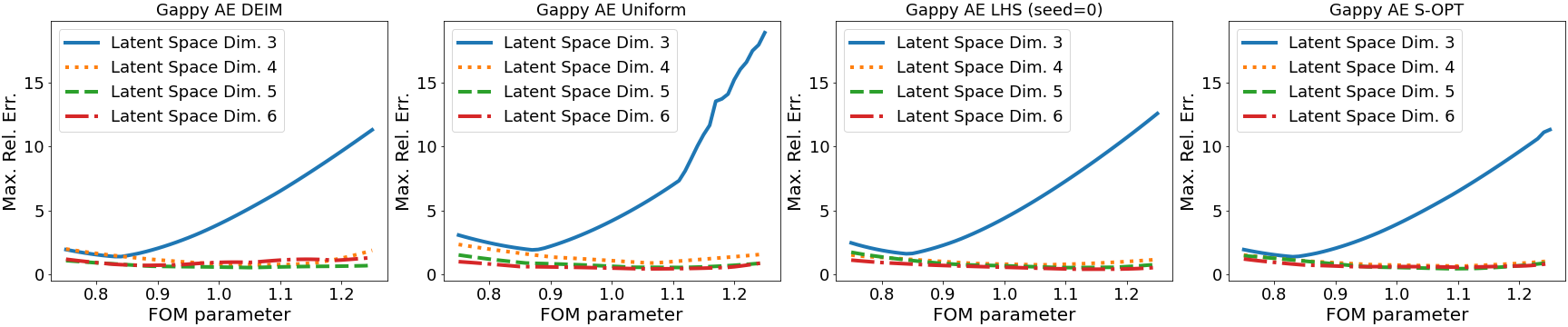}}
  \caption{Accuracy of Gappy AE for wave problem with noiseless measurements in the domain}
  \label{fg:ex23InnerGappyAESize}
\end{figure}

\begin{figure}[h]
  \centering
  \subfigure[Average Relative Error (\%)]{
  \includegraphics[width=0.9\textwidth]{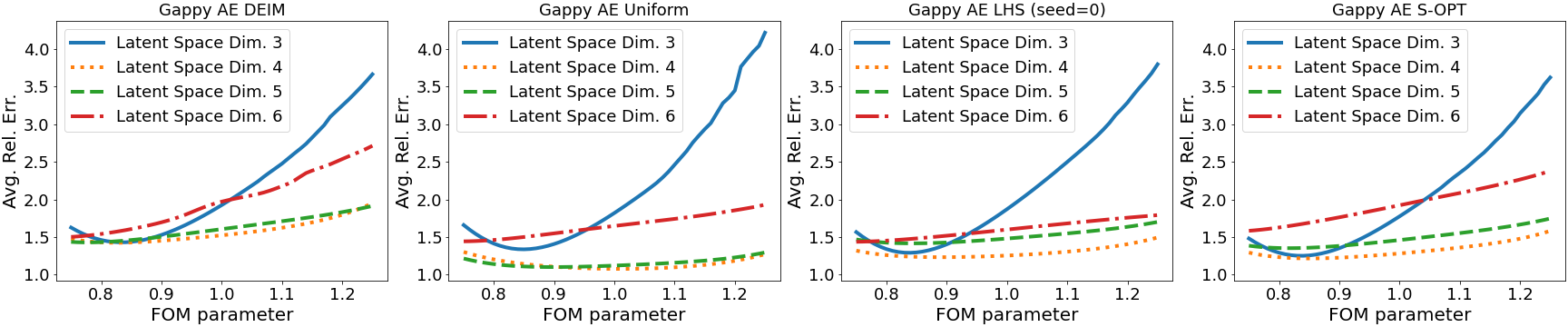}}
  \subfigure[Maximum Relative Error (\%)]{
  \includegraphics[width=0.9\textwidth]{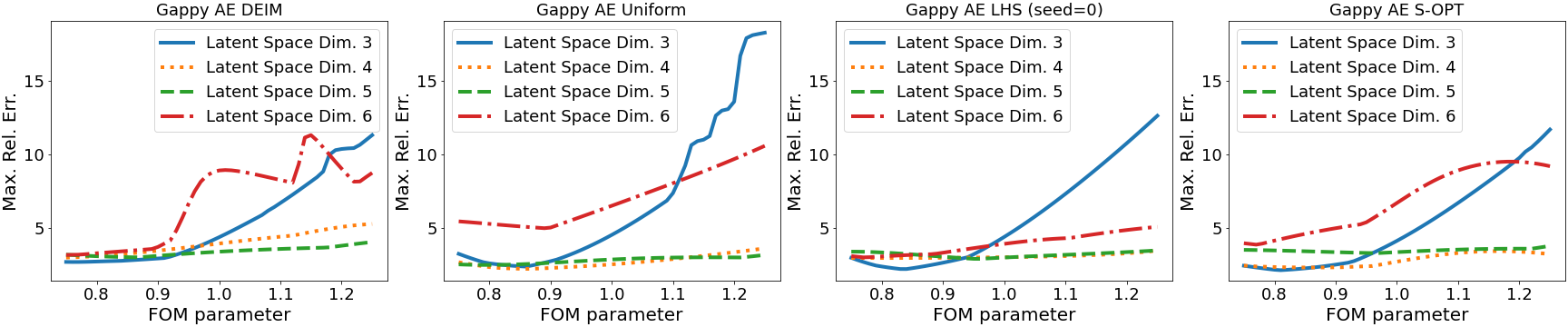}}
  \caption{Accuracy of Gappy AE for wave problem with noisy measurements in the domain}
  \label{fg:ex23InnerNoisyGappyAESize}
\end{figure}

The effects of varying the latent space dimension size are illustrated in Fig. \ref{fg:ex23InnerGappyAESize}. The relative errors decrease with increasing dimension except for the DEIM sampling case, where the relative error is not the lowest when the latent space dimension is 6. However, as shown in Fig. \ref{fg:ex23InnerNoisyGappyAESize}, the presence of noise in measurement data results in better accuracy for Gappy AE with a smaller latent space dimension (i.e., 4).
 \clearpage

 \clearpage

Next, comparisons of the Gappy AE performance for each sampling method in Figs. 22, 23, 24, and 25 of Appendix A.3.2 in \cite{appendix} reveal similar accuracy across sampling methods without noise and lowest relative error for the uniform sampling method with noise.

\begin{figure}[h]
    \centering
\subfigure[Average Relative Error (\%)]
{\includegraphics[width=0.45\textwidth]{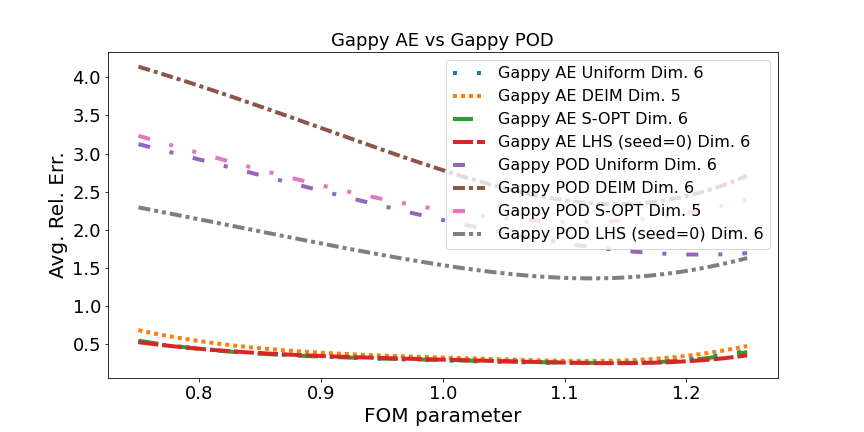}}
    \subfigure[Maximum Relative Error (\%)]
{\includegraphics[width=0.45\textwidth]{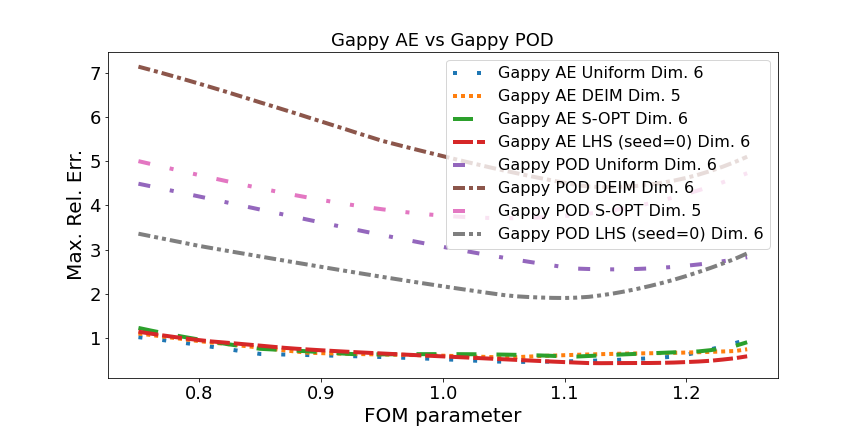}}
    \caption{Gappy AE vs. Gappy POD for wave problem with noiseless measurements in domain}
\label{fg:ex23InnerGappyAEvsGappyPOD}
\end{figure}

\begin{figure}[h]
    \centering
    \subfigure[]{
        \includegraphics[width=0.7\textwidth]{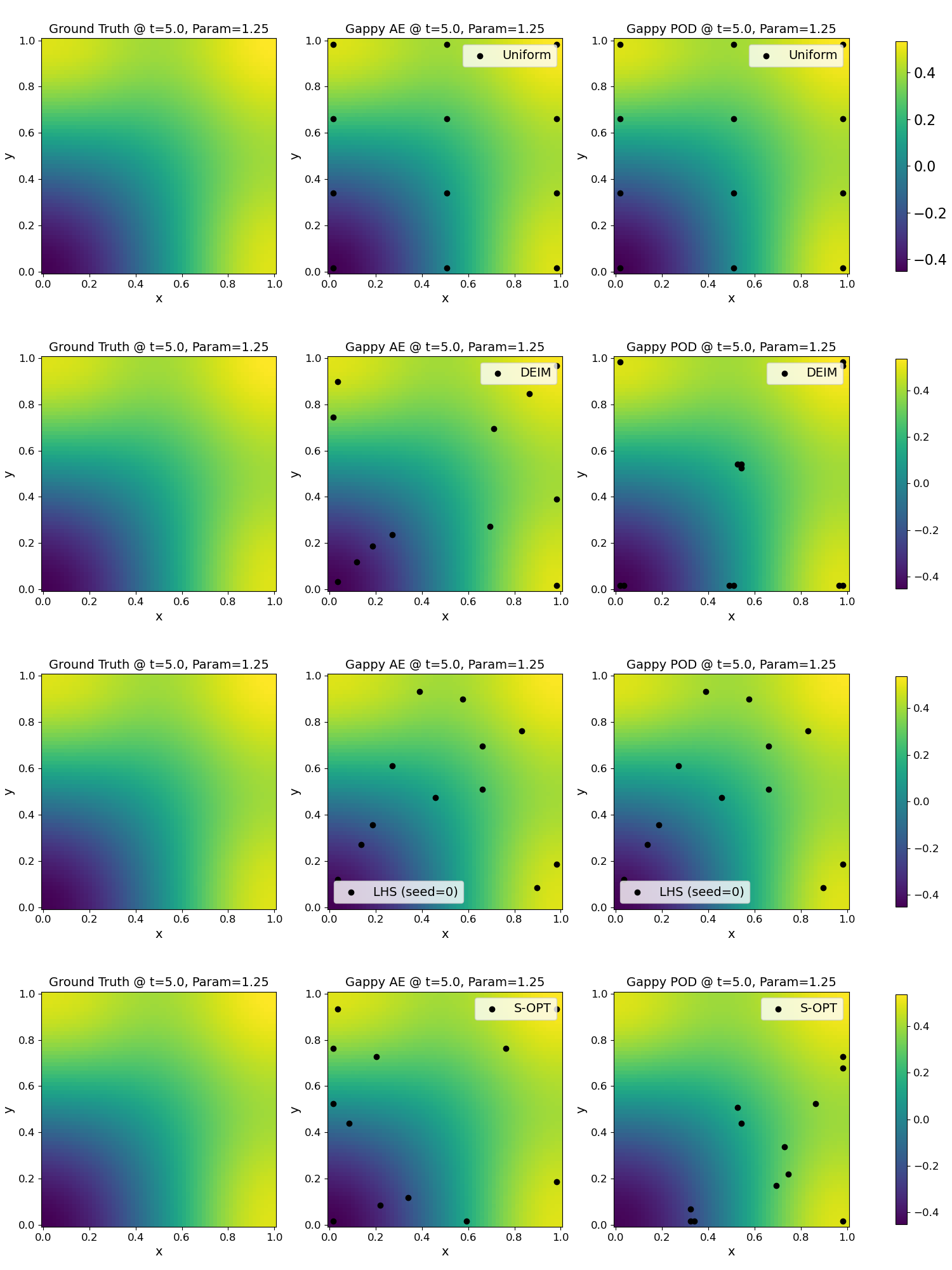}}
    \caption{Gappy AE vs. Gappy POD solutions for wave problem with noiseless measurements in the domain. Black dots denote sampling points}
\label{fg:ex23InnerGappyAEvsGappyPODSol}
\end{figure}

\begin{figure}[h]
    \centering
    \subfigure[Average Relative Error (\%)]
    {\includegraphics[width=0.45\textwidth]{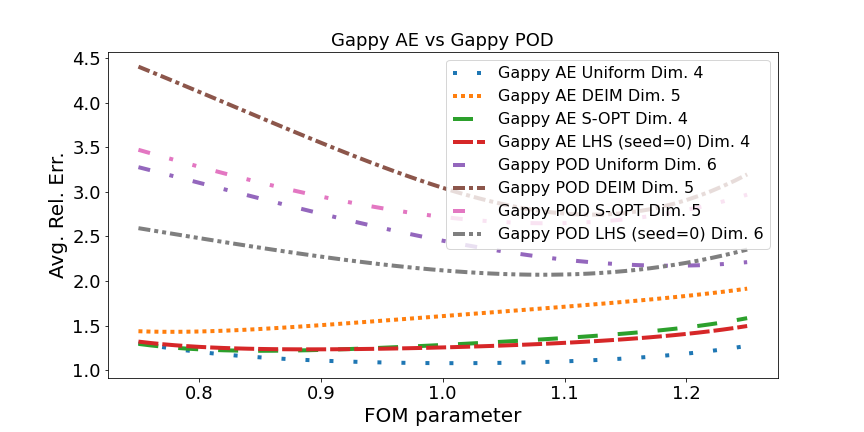}}
    \subfigure[Maximum Relative Error (\%)]
    {\includegraphics[width=0.45\textwidth]{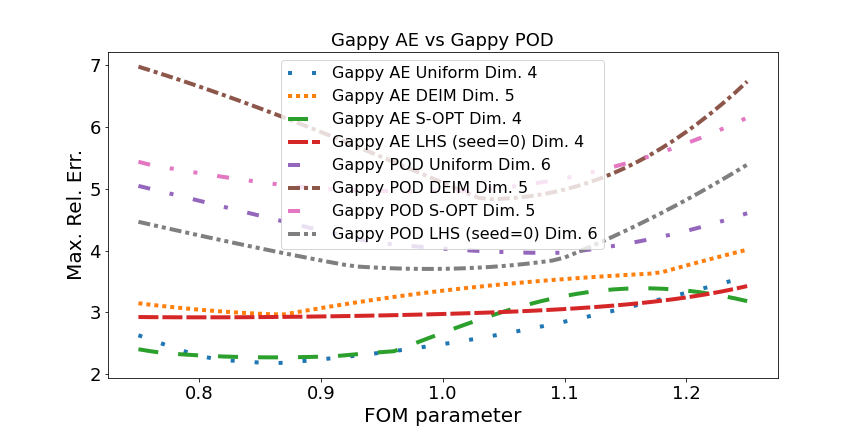}}
    \caption{Gappy AE vs. Gappy POD for wave problem with noisy measurements in the domain}
\label{fg:ex23InnerNoisyGappyAEvsGappyPOD}
\end{figure}

\begin{figure}[h]
    \centering
    \includegraphics[width=0.7\textwidth]{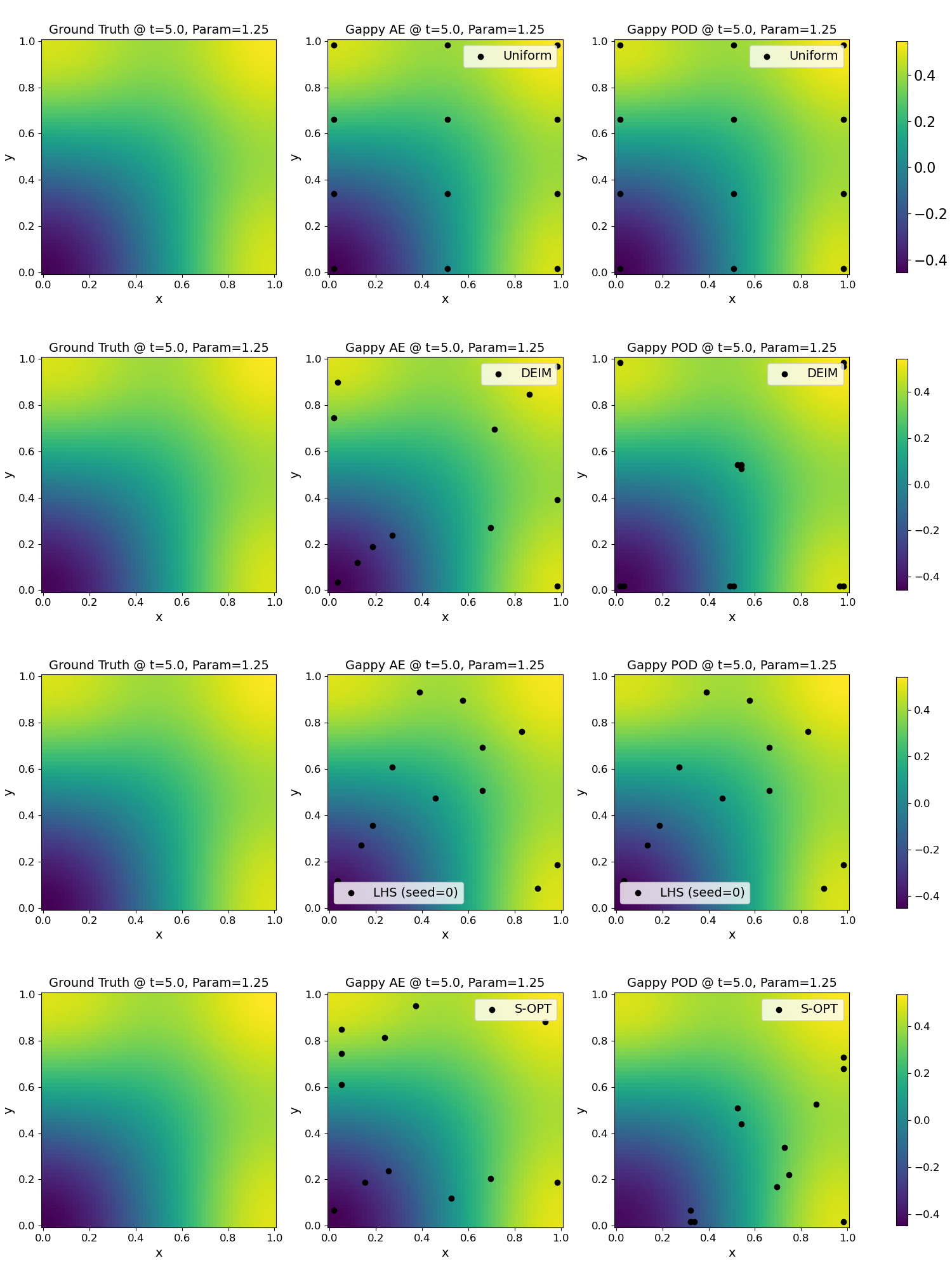}
    \caption{Gappy AE vs. Gappy POD solutions for wave problem with noisy measurements in the domain. Black dots denote sampling points}
\label{fg:ex23InnerNoisyGappyAEvsGappyPODSol}
\end{figure}

Finally, the Gappy AE and POD methods are compared in Figs. \ref{fg:ex23InnerGappyAEvsGappyPOD}, \ref{fg:ex23InnerGappyAEvsGappyPODSol}, \ref{fg:ex23InnerNoisyGappyAEvsGappyPOD}, and \ref{fg:ex23InnerNoisyGappyAEvsGappyPODSol}. Regardless of noise in the measurement data, the Gappy AE algorithm outperforms Gappy POD in accuracy.
 \clearpage

 \clearpage

 \clearpage

 \clearpage

\section{Discussion \& conclusion}\label{sec:discussion-conclusion}
We proposed the Gappy AE method for gappy data reconstruction and compared it with the popular Gappy POD method. In the Gappy AE method, data is reconstructed using a nonlinear manifold by training the auto-encoder. Unlike a linear subspace-based approach, such as the Gappy POD method, the Gappy AE shows $\mathcal{O}(10)$ to $\mathcal{O}(100)$ times less data reconstruction error when assuming noiseless measurement as shown in Table \ref{tb:GappyAEvsGappyPOD}.
\begin{table}[h]
\caption{Comparison of relative errors between Gappy AE and Gappy POD under noiseless measurement conditions. The best-performing cases from each numerical example are selected, and their relative errors are computed by averaging over FOM parameters.}\label{tb:GappyAEvsGappyPOD}
\centering
\begin{tabular}{|c|c|cc|}
\hline
\multirow{2}{*}{Problem Type}     & \multirow{2}{*}{Sensor Placement} & \multicolumn{2}{c|}{Relative Error (\%)}    \\ \cline{3-4} 
                                  &                                   & \multicolumn{1}{c|}{Gappy AE} & Gappy POD   \\ \hline
\multirow{2}{*}{Diffusion}        & Boundary                          & \multicolumn{1}{c|}{0.0896}    & 9.7730     \\ \cline{2-4} 
                                  & Domain                            & \multicolumn{1}{c|}{0.0753}    & 2.6964     \\ \hline
\multirow{1}{*}{Radial Advection} & Domain                            & \multicolumn{1}{c|}{0.0699}    & 7.8696     \\ \hline 
\multirow{2}{*}{Wave}             & Boundary                          & \multicolumn{1}{c|}{0.4102}    & 2.5992     \\ \cline{2-4} 
                                  & Domain                            & \multicolumn{1}{c|}{0.3272}    & 1.6759     \\ \hline
\end{tabular}
\end{table}
Even for diffusion dominant problems, the Gappy AE can outperform Gappy POD in terms of accuracy when the number of sample points is very sparse. However, we acknowledge that Gappy POD will in general outperform Gappy AE in terms of speed-up. The superior representability of the nonlinear manifold allows the Gappy AE to reconstruct gappy data when known (measured) data is sparse. This prominent feature is aligned with the digital twin application for physical data because, in reality, sensors are limited to measuring sparse physical field data, yet they want to estimate the entire field data. In addition to that, if there is training data available for autoencoders, the Gappy AE method can be used in any domains whether simulation is possible or not. In our numerical tests, the spatial dimension of reconstruction data is $3,600$ and the number of sample points is $12$. The speed of data reconstruction using the Gappy POD is approximately $30$ times faster than the Gappy AE. Although the Gappy AE takes more time to reconstruct data because of solving nonlinear problems in latent space dimension, its speed is sufficient for a $260$ Hz measurement frequency. 

As explained in Section \ref{sec:sampling}, DEIM and S-OPT algorithms are applied to the residual term to select sampling points for the Gappy AE. These algorithms find sample points to estimate the residual well, not the data to be reconstructed. Therefore, new algorithms to suggest sample points for the Gappy AE based on reconstruction error are needed.

The number of operations for the Gappy POD computation given by Eq. \eqref{eq:gappyPODcomputation} is $\mathcal{O}(\ndof\nbasisres)$, where $\ndof$ is the dimension of original data and $\nbasisres$ is the number of samples. $\basismatspace (\samplemat\basismatspace)^{\dagger}$ is precomputed once and $\samplemat \sol$ is not a matrix and vector multiplication but a sampling process. Thus, $\ndof \times \nbasisres$ shaped matrix and vector multiplication is done every measurement step, resulting in $\mathcal{O}(\ndof\nbasisres)$. For Gappy AE, generalized coordinates of latent space are computed first, and its computational cost is $\mathcal{O}(\nbasisres b f) + \mathcal{O}(f\nbasisres^2)$, where $b$ is the number of nodes in the hidden layer to compute single output nodes and $f$ is the latent space dimension. Then, decoder evaluation for output requires $\mathcal{O}(\ndof b)+\mathcal{O}(\ndof f\delta b)$ operations, where $\delta b$ is the amount by which the block of $b$ nodes shifts. Thus, we have $\mathcal{O}(\nbasisres bf)+\mathcal{O}(\nbasisres^2f)+\mathcal{O}(\ndof f\delta b)+\mathcal{O}(\ndof b)$ for computational cost of the Gappy AE for each measurement step. Details on counting operations are presented in \cite{kim2022fast}. Since we assume $\nbasisres << \ndof$, $\delta b < b << \ndof$, and $f << \ndof$, the computational cost of the Gappy AE increases linearly as the problem size, $\ndof$ increases, indicating that the Gappy AE is also suitable for a large problem if $\nbasisres$, $f$, $\delta b$, and $b$ are small.

In future work, we will investigate retraining the auto-encoder with noisy data to make the Gappy AE more reliable for noisy measurements. To be more specific, we plan to train two decoders. One is for noisy data and the other is for noiseless data. The decoder trained with noisy data can be used for finding $\redsolapprox$ in a latent space. Then, the decoder trained with noiseless data will be used to reconstruct data.

\section*{Acknowledgments}
This work was performed at the Korea Institute of Science and Technology and was supported by the Ministry of Trade, Industry, and Energy and the Korea Evaluation Institute of Industrial Technology research grant (20012462). This research was also supported by the Basic Science Research Program through the National Research Foundation of Korea, funded by the Ministry of Education (RS-2023-00272582). Y. Choi acknowledges support from the U.S. Department of Energy, Office of Science, Office of Advanced Scientific Computing Research, as part of the CHaRMNET Mathematical Multifaceted Integrated Capability Center (MMICC) program, under Award Number DE-SC0023164 at Lawrence Livermore National Laboratory. Lawrence Livermore National Laboratory is operated by Lawrence Livermore National Security, LLC, for the U.S. Department of Energy, National Nuclear Security Administration under Contract DE-AC52-07NA27344. IM release number: LLNL-JRNL-858295

\bibliographystyle{plain}
\bibliography{__references.bib}

\begin{thebibliography}{10}

\bibitem{mfem}
R.~Anderson, J.~Andrej, A.~Barker, J.~Bramwell, J.-S. Camier, J.~Cerveny, V.~Dobrev, Y.~Dudouit, A.~Fisher, Tz. Kolev, W.~Pazner, M.~Stowell, V.~Tomov, I.~Akkerman, J.~Dahm, D.~Medina, and S.~Zampini.
\newblock {MFEM}: A modular finite element methods library.
\newblock {\em Computers \& Mathematics with Applications}, 81:42--74, 2021.

\bibitem{berkooz1993proper}
Gal Berkooz, Philip Holmes, and John~L Lumley.
\newblock The proper orthogonal decomposition in the analysis of turbulent flows.
\newblock {\em Annual review of fluid mechanics}, 25(1):539--575, 1993.
\newblock doi: \url{https://doi.org/10.1146/annurev.fl.25.010193.002543}.

\bibitem{bonneville2024gplasdi}
Christophe Bonneville, Youngsoo Choi, Debojyoti Ghosh, and Jonathan~L Belof.
\newblock Gplasdi: Gaussian process-based interpretable latent space dynamics identification through deep autoencoder.
\newblock {\em Computer Methods in Applied Mechanics and Engineering}, 418:116535, 2024.
\newblock doi: \url{https://doi.org/10.1016/j.cma.2023.116535}.

\bibitem{cai2021physics_fluid}
Shengze Cai, Zhiping Mao, Zhicheng Wang, Minglang Yin, and George~Em Karniadakis.
\newblock Physics-informed neural networks (pinns) for fluid mechanics: A review.
\newblock {\em Acta Mechanica Sinica}, 37(12):1727--1738, 2021.
\newblock doi: \url{https://doi.org/10.1007/s10409-021-01148-1}.

\bibitem{cai2021physics_thermal}
Shengze Cai, Zhicheng Wang, Sifan Wang, Paris Perdikaris, and George~Em Karniadakis.
\newblock Physics-informed neural networks for heat transfer problems.
\newblock {\em Journal of Heat Transfer}, 143(6):060801, 2021.
\newblock doi: \url{https://doi.org/10.1115/1.4050542}.

\bibitem{carlberg2011efficient}
Kevin Carlberg, Charbel Bou-Mosleh, and Charbel Farhat.
\newblock Efficient non-linear model reduction via a least-squares petrov--galerkin projection and compressive tensor approximations.
\newblock {\em International Journal for numerical methods in engineering}, 86(2):155--181, 2011.
\newblock doi: \url{https://doi.org/10.1002/nme.3050}.

\bibitem{carlberg2018conservative}
Kevin Carlberg, Youngsoo Choi, and Syuzanna Sargsyan.
\newblock Conservative model reduction for finite-volume models.
\newblock {\em Journal of Computational Physics}, 371:280--314, 2018.
\newblock doi: \url{https://doi.org/10.1016/j.jcp.2018.05.019}.

\bibitem{carlberg2013gnat}
Kevin Carlberg, Charbel Farhat, Julien Cortial, and David Amsallem.
\newblock The gnat method for nonlinear model reduction: effective implementation and application to computational fluid dynamics and turbulent flows.
\newblock {\em Journal of Computational Physics}, 242:623--647, 2013.
\newblock doi: \url{https://doi.org/10.1016/j.jcp.2013.02.028}.

\bibitem{chaturantabut2010nonlinear}
Saifon Chaturantabut and Danny~C Sorensen.
\newblock Nonlinear model reduction via discrete empirical interpolation.
\newblock {\em SIAM Journal on Scientific Computing}, 32(5):2737--2764, 2010.
\newblock doi: \url{https://doi.org/10.1137/090766498}.

\bibitem{cheung2023local}
Siu~Wun Cheung, Youngsoo Choi, Dylan~Matthew Copeland, and Kevin Huynh.
\newblock Local lagrangian reduced-order modeling for the rayleigh-taylor instability by solution manifold decomposition.
\newblock {\em Journal of Computational Physics}, 472:111655, 2023.
\newblock doi: \url{https://doi.org/10.1016/j.jcp.2022.111655}.

\bibitem{doecode_24508}
Youngsoo Choi, William~J. Arrighi, Dylan~M. Copeland, Robert~W. Anderson, and Geoffrey~M. Oxberry.
\newblock librom.
\newblock [Computer Software] \url{https://doi.org/10.11578/dc.20190408.3}, oct 2019.

\bibitem{choi2020gradient}
Youngsoo Choi, Gabriele Boncoraglio, Spenser Anderson, David Amsallem, and Charbel Farhat.
\newblock Gradient-based constrained optimization using a database of linear reduced-order models.
\newblock {\em Journal of Computational Physics}, 423:109787, 2020.
\newblock doi: \url{https://doi.org/10.1016/j.jcp.2020.109787}.

\bibitem{choi2020space}
Youngsoo Choi, Peter Brown, Bill Arrighi, Robert Anderson, and Kevin Huynh.
\newblock Space--time reduced order model for large-scale linear dynamical systems with application to boltzmann transport problems.
\newblock {\em Journal of Computational Physics}, P109845, 2020.
\newblock doi: \url{https://doi.org/10.1016/j.jcp.2020.109845}.

\bibitem{choi2019space}
Youngsoo Choi and Kevin Carlberg.
\newblock Space--time least-squares petrov--galerkin projection for nonlinear model reduction.
\newblock {\em SIAM Journal on Scientific Computing}, 41(1):A26--A58, 2019.
\newblock doi: \url{https://doi.org/10.1137/17M1120531}.

\bibitem{choi2020sns}
Youngsoo Choi, Deshawn Coombs, and Robert Anderson.
\newblock Sns: A solution-based nonlinear subspace method for time-dependent model order reduction.
\newblock {\em SIAM Journal on Scientific Computing}, 42(2):A1116--A1146, 2020.
\newblock doi: \url{https://doi.org/10.1137/19M1242963}.

\bibitem{choi2019accelerating}
Youngsoo Choi, Geoffrey Oxberry, Daniel White, and Trenton Kirchdoerfer.
\newblock Accelerating design optimization using reduced order models, 2019.
\newblock doi: \url{https://doi.org/10.48550/arXiv.1909.11320}.

\bibitem{cocola2023hyper}
Jorio Cocola, John Tencer, Francesco Rizzi, Eric Parish, and Patrick Blonigan.
\newblock Hyper-reduced autoencoders for efficient and accurate nonlinear model reductions.
\newblock {\em arXiv preprint arXiv:2303.09630}, 2023.
\newblock doi: \url{https://doi.org/10.48550/arXiv.2303.09630}.

\bibitem{copeland2022reduced}
Dylan~Matthew Copeland, Siu~Wun Cheung, Kevin Huynh, and Youngsoo Choi.
\newblock Reduced order models for lagrangian hydrodynamics.
\newblock {\em Computer Methods in Applied Mechanics and Engineering}, 388:114259, 2022.
\newblock doi: \url{https://doi.org/10.1016/j.cma.2021.114259}.

\bibitem{cuomo2022scientific}
Salvatore Cuomo, Vincenzo~Schiano Di~Cola, Fabio Giampaolo, Gianluigi Rozza, Maziar Raissi, and Francesco Piccialli.
\newblock Scientific machine learning through physics--informed neural networks: Where we are and what’s next.
\newblock {\em Journal of Scientific Computing}, 92(3):88, 2022.
\newblock doi: \url{https://doi.org/10.1007/s10915-022-01939-z}.

\bibitem{diaz2023fast}
Alejandro~N. Diaz, Youngsoo Choi, and Matthias Heinkenschloss.
\newblock A fast and accurate domain-decomposition nonlinear manifold reduced order model, 2023.
\newblock doi: \url{https://doi.org/10.48550/arXiv.2305.15163}.

\bibitem{drmac2016new}
Zlatko Drmac and Serkan Gugercin.
\newblock A new selection operator for the discrete empirical interpolation method---improved a priori error bound and extensions.
\newblock {\em SIAM Journal on Scientific Computing}, 38(2):A631--A648, 2016.
\newblock doi: \url{https://doi.org/10.1137/15M1019271}.

\bibitem{drmac2018discrete}
Zlatko Drmac and Arvind~Krishna Saibaba.
\newblock The discrete empirical interpolation method: Canonical structure and formulation in weighted inner product spaces.
\newblock {\em SIAM Journal on Matrix Analysis and Applications}, 39(3):1152--1180, 2018.
\newblock doi: \url{https://doi.org/10.1137/17M1129635}.

\bibitem{everson1995karhunen}
Richard Everson and Lawrence Sirovich.
\newblock Karhunen--loeve procedure for gappy data.
\newblock {\em JOSA A}, 12(8):1657--1664, 1995.
\newblock doi: \url{https://doi.org/10.1364/JOSAA.12.001657}.

\bibitem{fries2022lasdi}
William~D Fries, Xiaolong He, and Youngsoo Choi.
\newblock Lasdi: Parametric latent space dynamics identification.
\newblock {\em Computer Methods in Applied Mechanics and Engineering}, 399:115436, 2022.
\newblock doi: \url{https://doi.org/10.1016/j.cma.2022.115436}.

\bibitem{greif2019decay}
Constantin Greif and Karsten Urban.
\newblock Decay of the kolmogorov n-width for wave problems.
\newblock {\em Applied Mathematics Letters}, 96:216--222, 2019.
\newblock doi: \url{https://doi.org/10.1016/j.aml.2019.05.013}.

\bibitem{gundersen2021semi}
Kristian Gundersen, Anna Oleynik, Nello Blaser, and Guttorm Alendal.
\newblock Semi-conditional variational auto-encoder for flow reconstruction and uncertainty quantification from limited observations.
\newblock {\em Physics of Fluids}, 33(1), 2021.
\newblock doi: \url{https://doi.org/10.1063/5.0025779}.

\bibitem{he2023glasdi}
Xiaolong He, Youngsoo Choi, William~D. Fries, Jonathan~L. Belof, and Jiun-Shyan Chen.
\newblock glasdi: Parametric physics-informed greedy latent space dynamics identification.
\newblock {\em Journal of Computational Physics}, 489:112267, 2023.
\newblock doi: \url{https://doi.org/10.1016/j.jcp.2023.112267}.

\bibitem{hinze2005proper}
Michael Hinze and Stefan Volkwein.
\newblock Proper orthogonal decomposition surrogate models for nonlinear dynamical systems: Error estimates and suboptimal control.
\newblock In {\em Dimension reduction of large-scale systems}, pages 261--306. Springer, 2005.
\newblock doi: \url{https://doi.org/10.1007/3-540-27909-1_10}.

\bibitem{hotelling1933analysis}
Harold Hotelling.
\newblock Analysis of a complex of statistical variables into principal components.
\newblock {\em Journal of educational psychology}, 24(6):417, 1933.
\newblock doi: \url{https://doi.org/10.1037/h0071325}.

\bibitem{kim2020efficient}
Youngkyu Kim, Youngsoo Choi, David Widemann, and Tarek Zohdi.
\newblock Efficient nonlinear manifold reduced order model, 2020.
\newblock doi: \url{https://doi.org/10.48550/arXiv.2011.07727}.

\bibitem{kim2022fast}
Youngkyu Kim, Youngsoo Choi, David Widemann, and Tarek Zohdi.
\newblock A fast and accurate physics-informed neural network reduced order model with shallow masked autoencoder.
\newblock {\em Journal of Computational Physics}, 451:110841, 2022.
\newblock doi: \url{https://doi.org/10.1016/j.jcp.2021.110841}.

\bibitem{appendix}
Youngkyu Kim, Youngsoo Choi, and Byounghuyn Yoo.
\newblock Appendix: Gappy data reconstruction using unsupervised learning for digital twin.
\newblock \url{https://doi.org/10.17632/2xtvw6z2n9.1}, 2023.

\bibitem{kim2021efficient}
Youngkyu Kim, Karen Wang, and Youngsoo Choi.
\newblock Efficient space--time reduced order model for linear dynamical systems in python using less than 120 lines of code.
\newblock {\em Mathematics}, 9(14):1690, 2021.
\newblock doi: \url{https://doi.org/10.3390/math9141690}.

\bibitem{kunisch2002galerkin}
Karl Kunisch and Stefan Volkwein.
\newblock Galerkin proper orthogonal decomposition methods for a general equation in fluid dynamics.
\newblock {\em SIAM Journal on Numerical analysis}, 40(2):492--515, 2002.
\newblock doi: \url{https://doi.org/10.1137/S0036142900382612}.

\bibitem{kutz2016dynamic}
J~Nathan Kutz, Steven~L Brunton, Bingni~W Brunton, and Joshua~L Proctor.
\newblock {\em Dynamic mode decomposition: data-driven modeling of complex systems}.
\newblock SIAM, 2016.
\newblock doi: \url{https://doi.org/10.1137/1.9781611974508}.

\bibitem{lauzon2022s}
Jessica~T. Lauzon, Siu~Wun Cheung, Yeonjong Shin, Youngsoo Choi, Dylan~Matthew Copeland, and Kevin Huynh.
\newblock S-opt: A points selection algorithm for hyper-reduction in reduced order models, 2022.
\newblock doi: \url{https://doi.org/10.48550/arXiv.2203.16494}.

\bibitem{lee2020model}
Kookjin Lee and Kevin~T Carlberg.
\newblock Model reduction of dynamical systems on nonlinear manifolds using deep convolutional autoencoders.
\newblock {\em Journal of Computational Physics}, 404:108973, 2020.
\newblock doi: \url{https://doi.org/10.1016/j.jcp.2019.108973}.

\bibitem{loeve1955}
Michel Loeve.
\newblock {\em Probability Theory}.
\newblock D. Van Nostrand, New York, 1955.

\bibitem{mcbane2021component}
Sean McBane and Youngsoo Choi.
\newblock Component-wise reduced order model lattice-type structure design.
\newblock {\em Computer methods in applied mechanics and engineering}, 381:113813, 2021.
\newblock doi: \url{https://doi.org/10.1016/j.cma.2021.113813}.

\bibitem{mcbane2022stress}
Sean McBane, Youngsoo Choi, and Karen Willcox.
\newblock Stress-constrained topology optimization of lattice-like structures using component-wise reduced order models.
\newblock {\em Computer Methods in Applied Mechanics and Engineering}, 400:115525, 2022.
\newblock doi: \url{https://doi.org/10.1016/j.cma.2022.115525}.

\bibitem{mckay2000comparison}
M.~D. Mckay, R.~J. Beckman, and W.~J. Conover.
\newblock A comparison of three methods for selecting values of input variables in the analysis of output from a computer code.
\newblock {\em Technometrics}, 42(1):55--61, 2000.
\newblock doi: \url{https://doi.org/10.2307/1271432}.

\bibitem{mfem-web}
{MFEM}: Modular finite element methods {[Software]}.
\newblock \url{https://mfem.org}.

\bibitem{nair2020leveraging}
Nirmal~J Nair and Andres Goza.
\newblock Leveraging reduced-order models for state estimation using deep learning.
\newblock {\em Journal of Fluid Mechanics}, 897:R1, 2020.
\newblock doi: \url{https://doi.org/10.1017/jfm.2020.409}.

\bibitem{ohlberger2015reduced}
Mario Ohlberger and Stephan Rave.
\newblock Reduced basis methods: Success, limitations and future challenges, 2016.
\newblock doi: \url{https://doi.org/10.48550/arXiv.1511.02021}.

\bibitem{raissi2019physics}
Maziar Raissi, Paris Perdikaris, and George~E Karniadakis.
\newblock Physics-informed neural networks: A deep learning framework for solving forward and inverse problems involving nonlinear partial differential equations.
\newblock {\em Journal of Computational physics}, 378:686--707, 2019.
\newblock doi: \url{https://doi.org/10.1016/j.jcp.2018.10.045}.

\bibitem{romor2023non}
Francesco Romor, Giovanni Stabile, and Gianluigi Rozza.
\newblock Non-linear manifold reduced-order models with convolutional autoencoders and reduced over-collocation method.
\newblock {\em Journal of Scientific Computing}, 94(3):74, 2023.
\newblock doi: \url{https://doi.org/10.1007/s10915-023-02128-2}.

\bibitem{schmid2010dynamic}
Peter~J Schmid.
\newblock Dynamic mode decomposition of numerical and experimental data.
\newblock {\em Journal of fluid mechanics}, 656:5--28, 2010.
\newblock doi: \url{https://doi.org/10.1017/S0022112010001217}.

\bibitem{shin2016nonadaptive}
Yeonjong Shin and Dongbin Xiu.
\newblock Nonadaptive quasi-optimal points selection for least squares linear regression.
\newblock {\em SIAM Journal on Scientific Computing}, 38(1):A385--A411, 2016.
\newblock doi: \url{https://doi.org/10.1137/15M1015868}.

\bibitem{taddei2021space}
Tommaso Taddei and Lei Zhang.
\newblock Space-time registration-based model reduction of parameterized one-dimensional hyperbolic pdes.
\newblock {\em ESAIM: Mathematical Modelling and Numerical Analysis}, 55(1):99--130, 2021.
\newblock doi: \url{https://doi.org/10.1051/m2an/2020073}.

\bibitem{willcox2006unsteady}
Karen Willcox.
\newblock Unsteady flow sensing and estimation via the gappy proper orthogonal decomposition.
\newblock {\em Computers \& fluids}, 35(2):208--226, 2006.
\newblock doi: \url{https://doi.org/10.1016/j.compfluid.2004.11.006}.

\end{thebibliography}

\end{document}